\pdfoutput=1 
\documentclass[%
 twocolumn, 
 aps,
 superscriptaddres,
 groupedaddress,
 10pt
]{revtex4-2} 

\usepackage[english]{babel}
\usepackage{mathtools}
\usepackage{amsmath, amssymb}
\usepackage[utf8]{inputenc}
\usepackage{hyperref}
\usepackage[up]{caption}
\usepackage{subcaption}
\usepackage{dsfont}
\usepackage{graphicx}

\usepackage[version=4]{mhchem}
\usepackage[T1]{fontenc}
\usepackage{lmodern}
\usepackage{braket} 
\usepackage{color}
\usepackage[dvipsnames]{xcolor}

\makeatletter
\def\maketitle{
\@author@finish
\title@column\titleblock@produce
\suppressfloats[t]}
\makeatother

\begin{document}

\preprint{APS/123-QED}

\title{Fractional quantum Hall states with variational Projected Entangled-Pair States:\\
a study of the bosonic Harper-Hofstadter model
}

\author{Erik Lennart Weerda}
\email[corresponding author: ]{weerda@thp.uni-koeln.de}

\affiliation{Institute for Theoretical Physics, University of Cologne, D-50937 K\"oln, Germany}

\author{Matteo Rizzi}%
\affiliation{Institute for Theoretical Physics, University of Cologne, D-50937 K\"oln, Germany}
\affiliation{%
 Forschungszentrum J\"ulich  GmbH, Institute of Quantum Control, Peter Gr\"unberg Institut (PGI-8), 52425 J\"ulich, Germany
}%

\date{\today}

\begin{abstract}
An important class of model Hamiltonians for investigation of topological phases of matter consists of mobile, interacting particles on a lattice subject to a semi-classical gauge field, as exemplified by the bosonic Harper-Hofstadter model. A unique method for investigations of two-dimensional quantum systems are the infinite projected-entangled pair states (iPEPS), as they avoid spurious finite size effects that can alter the phase structure. 
However, due to no-go theorems in related cases this was often conjectured to be impossible in the past. In this letter, we show that upon variational optimization the infinite projected-entangled pair states can be used to this end, by identifying fractional Hall states in the bosonic Harper-Hofstadter model. The obtained states are characterized by showing exponential decay of bulk correlations, as dictated by a bulk gap, as well as chiral edge modes via the entanglement spectrum.
\end{abstract}

\maketitle

\textit{Introduction.}
$-$The physics of interacting topological phases of matter, which includes phenomena like fractionally charged quasiparticles, exotic exchange statistics and long-range (topological) entanglement has led to the formulation of many new concepts in physics, most notably topological order \cite{Wen2017ZOO}. 
Additionally, using these phenomena offers a promising route for robust quantum information technology \cite{kitaev2003fault, DasSarma2008TQC}.
The paradigmatic family of quantum states for these phases of matter are the fractional quantum Hall states \cite{Stromer82FQH, Laughlin83FQH} and their lattice generalization as fractional Chern insulators \cite{Parameswaran2013}. 
Given this, an enormous amount of theoretical effort, driven by the fundamental interest in the fractional quantum Hall states has been put forward since their discovery and continues to this day.
Moreover, within the last years advances, in capabilities of experiments on cold atoms, like the possibility to include artificial gauge fields have put the extremely rich physics of the fractional Hall states into the grasp of current technologies  \cite{Aidelsburger2011, Aidelsburger2013, Ketterle2013, Bloch2014chrialcurrent, Greiner2017microscopy}. Recently, a crucial first experimental step was made with the realization of a Laughlin state of two atoms in such an experiment \cite{Greiner2023FQHrealization}.
A major part of the quest for the realization of the fractional quantum Hall physics in well controlled synthetic quantum matter experiments is the identification of specific Hamiltonians where these phases arise.
As these states generically emerge from an intricate interplay of magnetic fields and interactions between the particles, the candidate Hamiltonians are challenging for state of the art numerical methods to treat.
Within the class of tensor network methods for such numerical studies, the infinite Projected Entangled-Pair State (iPEPS) \cite{verstraete2004PEPS} framework has unique advantages for this task. 
As it operates directly in the thermodynamic limit in two dimensions it can eliminate finite size effects that can significantly bias other related methods, e.g. cylinder-MPS, leading to possibly altered phase boundaries or even artificial phases that are non-existent in the two dimensional limit.
Thus, the iPEPS ansatz constitutes an important tool for the determination of the phase structure of many-body Hamiltonians. 
However, doubts had been raised about the general applicability of the PEPS ansatz for the numerical simulation of gapped chiral states because of a no-go theorem for the case for free fermions \cite{Read2015NOGO}. In a recent work this issue was studied more closely from a numerical perspective \cite{Hasik2022CSL}. 
It was shown that for a truncated parent Hamiltonian of a chiral spin liquid \cite{Nielson2012CSLparent1, Nielsen2013CSLparentlocal} the iPEPS ansatz was indeed successful in finding and representing this ground state numerically.\\
In this work, we take a further step in utility of the iPEPS framework,
by treating for the first time mobile bosons in an external gauge field, showing that chiral gapped topological phases arise.
We are able to treat flux values that are in the range of today's experimental technology. This setting necessitates the use of large unit cells in the iPEPS ansatz due to the magnetic unit cell over which the Hamiltonian terms vary due to the external gauge field. 
Crucial for this investigation is the fact that we make use of a fully variational ground state search, as attempts with simple imaginary time evolution turn out to yield inconclusive results.
By scanning through different parameter values of the Harper-Hofstadter Hamiltonian we are able to find incompressible plateaus at the filling fraction expected for bosonic fractional Hall states. We focus on the paradigmic case of the Laughlin state in the case of hard-core bosons, but show that also for soft-core bosons and higher fillings we can find incompressible plateaus corresponding to more exotic fractional Hall states.   
The chiral nature of these states is verified with the use of the entanglement spectrum. We thus for the first time represent bosonic fractional quantum Hall states with PEPS. \\
\textit{Model and Methods.}
$-$ We study a two dimensional system of bosonic particles on a lattice, which are interacting on-site and are subject to a magnetic flux $\phi$ through each plaquette of the lattice. This situation is described by a interacting Harper-Hofstadter Hamiltonian
\begin{equation}
    \hat{H} = \sum_{\langle ij \rangle} (e^{i A_{ij}} \hat{a}_i^{\dagger} \hat{a}_j + h.c.) - \mu \sum_i \hat{n}_i + \frac{U}{2}\sum_i \hat{n}_i(\hat{n}_i -1),
    \label{eq: Hamiltonian}
\end{equation}
in which the hopping amplitude $t$ is dressed by the semi-classical gauge field. The gauge field is chosen such that it produces the desired magnetic flux per plaquette, $\phi = \sum_{\square} A_{ij}$. 
This Hamiltonian can be realized in cold atom experiments \cite{Aidelsburger2013, Ketterle2013, Greiner2017microscopy}, and was the basis for the recent preparation of a two-body fractional quantum Hall state \cite{Greiner2023FQHrealization}. In the limit of large fillings it also describes the physics of Josephson Junction Arrays \cite{fazio2001JJA}.\\
For low densities and small values of $\phi$, where the magnetic length $\ell_B = 1/\sqrt{\phi}$ is much larger than the lattice spacing, we approach the continuum setting, in which the presence of bosonic fractional Hall states is well established \cite{Read1999}.
However, at large flux values at which current cold atom experiments can operate and lattice effects play a central role, presence of truly lattice versions of the fractional quantum Hall states in the phase diagram remains a question of ongoing research.
Studies based on exact diagonalization \cite{Lukin2005, Lukin2007, Moller2009BIQH, Rizzi2010, Moller2012, He2015BIQH, Sterdyniak2015BIQH} as well as (infinite) density matrix renormalization group ((i)DMRG) \cite{Vishvanath2017, Jaksch2019DMRG, Grusdt2021DMRG, Boesl2022} and tree-tensor networks (TTN) \cite{Rizzi2017TTN-FQH, Rizzi2020} have suggested a multitude of fractional quantum Hall states in such systems. 
Even though these methods provide inside into the possible phases of the Hamiltonian in the two dimensional thermodynamic limit, it is widely acknowledged that they suffer from strong finite size effects. In particular for elongated cylinder geometries, as usually employed by the (i)DMRG methods of treating two dimensional Hamiltonians, these strong finite size effects lead to an overestimation of the stability of gapped phases as well as the emergence of charge density wave orders, breaking translation invariance, as discussed in  \cite{Grusdt2021DMRG, Boesl2022}.
Generically, it is not clear what the influence of finite size effects is on all the possible low-energy phases close to the true ground state is. Thus if these phases are in close competition, the phase with energetically most favorable finite size effect can be accidentally identified as the ground state. 
The circumstances where these considerations can become crucial include quite important models like the doped Hubbard model or models relevant for the search for spin liquids in materials just to name a few. \\
Here, we thus aim to provide important clarification on the situation by showing that the iPEPS ansatz, using variational optimization, is capable of tackling such problems directly in the thermodynamic limit, and hence without spurious finite size effects.
The iPEPS ansatz \cite{verstraete2004PEPS, Jordan2008iPEPS} for the many-body wavefunction 
used in this investigation is parameterized by a set of tensors $(A_{1,1} , \dots A_{m,n})$ that form a $m \times n$ unit cell (illustrated with the grey line) which is periodically repeated on the infinite lattice: 
\begin{equation}
\begin{split}
    \ket{\Psi(A_{1,1}, \dots, A_{m,n})} &= \sum_{\{n_i\}} C_{\{n_i\}}(A_{1,1}, \dots, A_{m,n}) \ket{\{n_i\}}\\
    C_{\{n_i\}}(A_{1,1}, \dots, A_{m,n}) &= \begin{gathered}
\includegraphics[height=3cm]{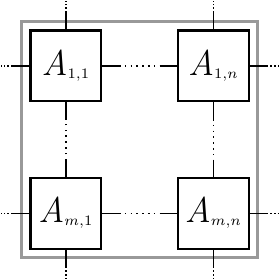}
\end{gathered},
\end{split}
\end{equation}
The individual tensors $A_{i,j}$ are rank 5 tensors with 4 virtual indices of bulk (bond-)dimension $\chi_B$ as well as a physical index whose dimension, $d$, is that of the local Hilbert space of the physical system.
In order to contract this infinite tensor network for the calculation of norms and expectation values, we make use of the corner transfer matrix renormalization group (CTMRG) method \cite{Baxter1968CTMRG1, Baxter78CTMRG2, Nishino96CTMRG, Nishino97CTMRG, Orus2009CTMRG, Corboz2010, Corboz2011, Corboz2014}. 
The CTMRG is a power method that generates effective environments approximating the contraction of subsections of the infitite tensor network. 
The accuracy of this approximation is controlled by the bond dimension of the environment tensors, which we call the \textit{environment bond dimension} $\chi_E$. 
Beyond the calculation of expectation values, the environment tensors generated with the CTMRG method can also be utilized in other ways, e.g. in the evaluation of the entanglement spectrum \cite{Li&Haldane2008, Haegeman2017Medley}, cf. supplementary material \cite{SupplementaryMaterial}. \\
In order to allow us to \textit{for the first time} treat more complicated Hamiltonians with a semi-classical gauge field and (large) magnetic unit cell and hosting chiral topological ground states with iPEPS we need to perform a variational ground state search.
Within the ansatz-class of iPEPS of a fixed unit cell size and bond dimension $\chi_B$, the variational state-optimization \cite{Corboz2016Variational, Vanderstraeten2016Variational} using automatic differentiation \cite{Liao2019AD}, implemented as explained in \cite{varipeps}, allows us to find the optimal approximation of the ground state.
This variational framework for iPEPS has been advantageous in treating e.g. frustrated spin system including recently spin liquids \cite{Hasik2021, Hasik2022, Hasik2022CSL, Corboz2016Variational}. 
For Hamiltonians, as in eq.~\eqref{eq: Hamiltonian} a treatment using the alternative methods based on imaginary-time evolution (simple update) did not yield useful results. This further illustrates the importance of the framework of variational optimization of the iPEPS ansatz.

\textit{Results.}
$-$
\begin{figure}[t]
    \centering
    \includegraphics[width=0.4\textwidth]{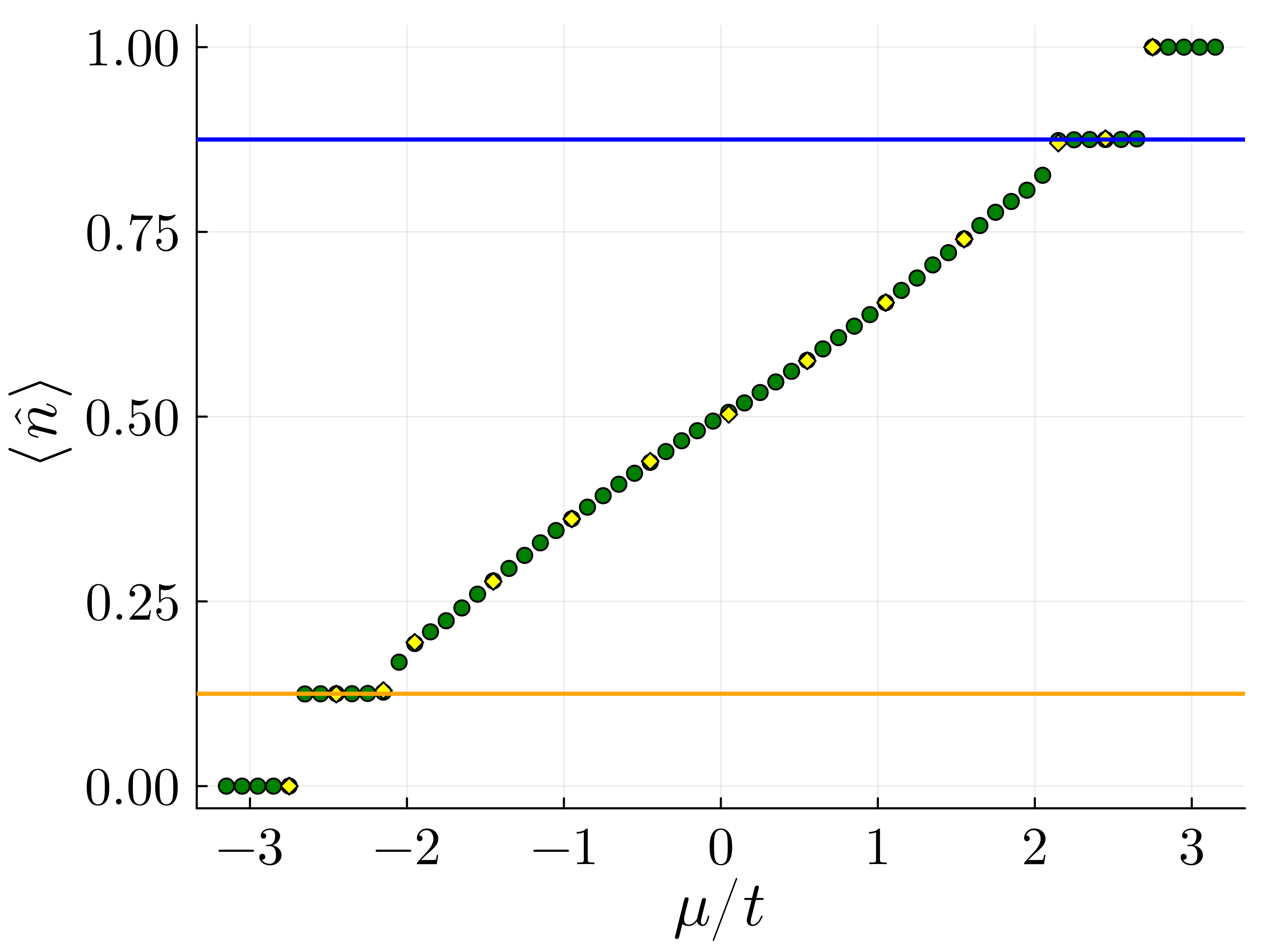}
    \caption{\raggedright \small Average occupation per site $\langle \hat{n}\rangle$ as a function of $\mu / t$ for hard-core bosons with bond dimensions $\chi_B = 4$ (\textit{green circles}) and $\chi_B=5$ (\textit{yellow diamonds}). Note that the two incompressible plateaus are found at the filling expected for the $\nu = 1/2$ Laughlin state and its hole analog. The regime given by the fractional Hall plateaus makes up more than 20\% of the parameter-space in between the empty and completely filled state.}
    \label{fig:occ_hardcore}
\end{figure}
\begin{figure}[t]
    \centering
    \begin{subfigure}{.4\textwidth}
        \includegraphics[width=\textwidth]{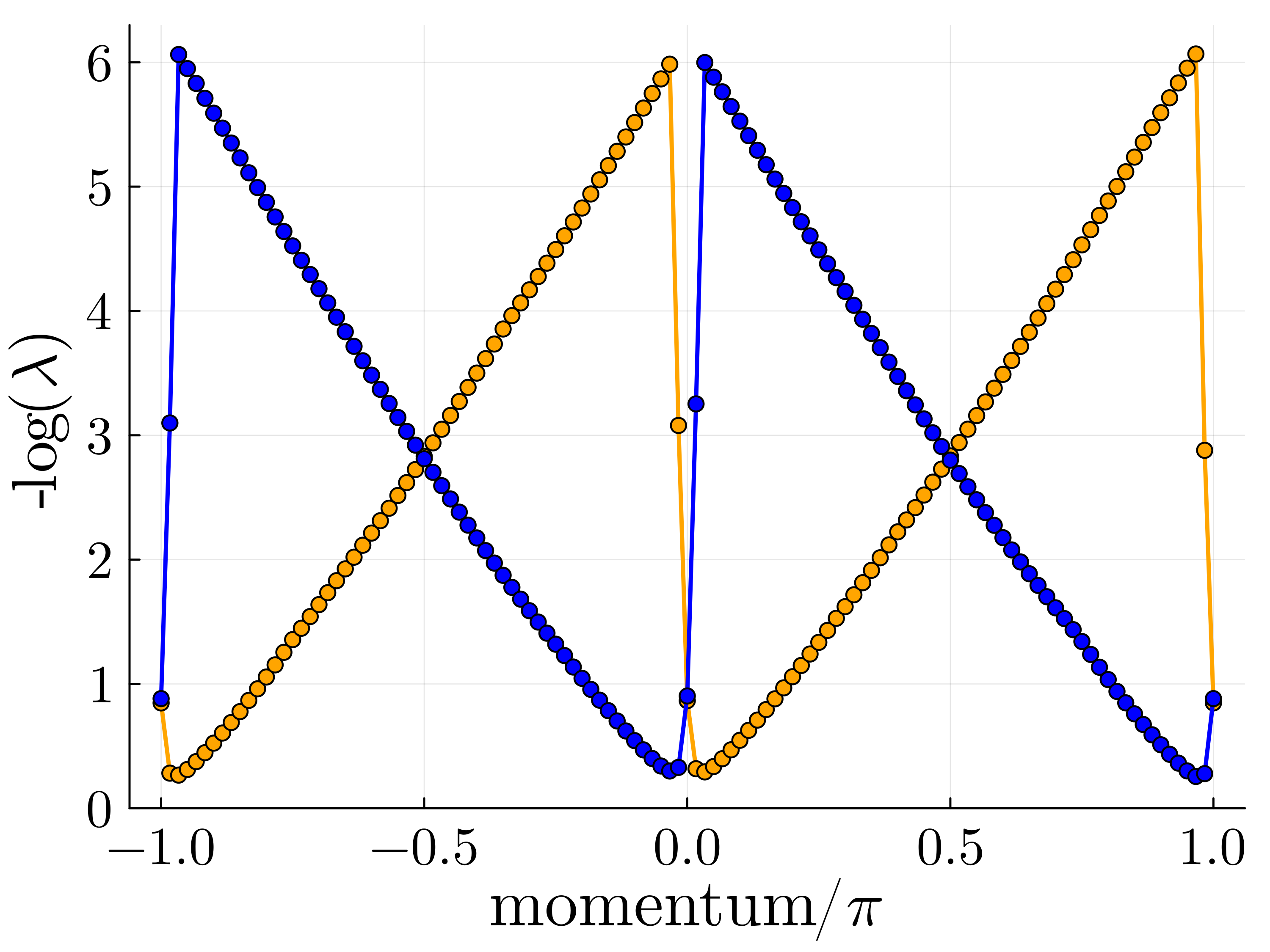}
        \caption{}
        \captionsetup{justification=centering}
        \label{fig:ES_combi}
    \end{subfigure}
    \begin{subfigure}{.2\textwidth}
        \includegraphics[width=\textwidth]{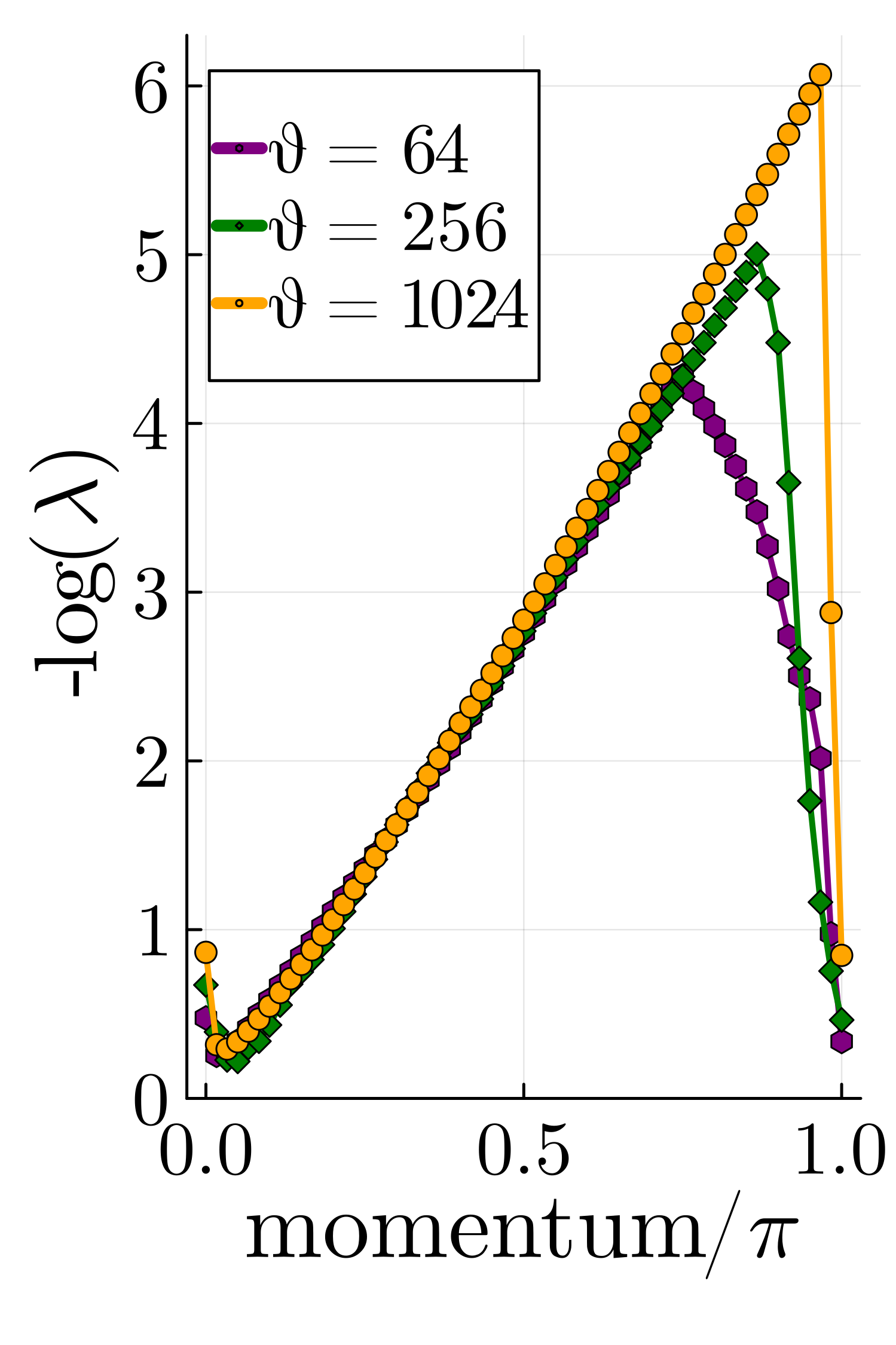}
        \caption{}
        \label{fig:ES parameters}
    \end{subfigure}
    \begin{subfigure}{.2\textwidth}
        \includegraphics[width=\textwidth]{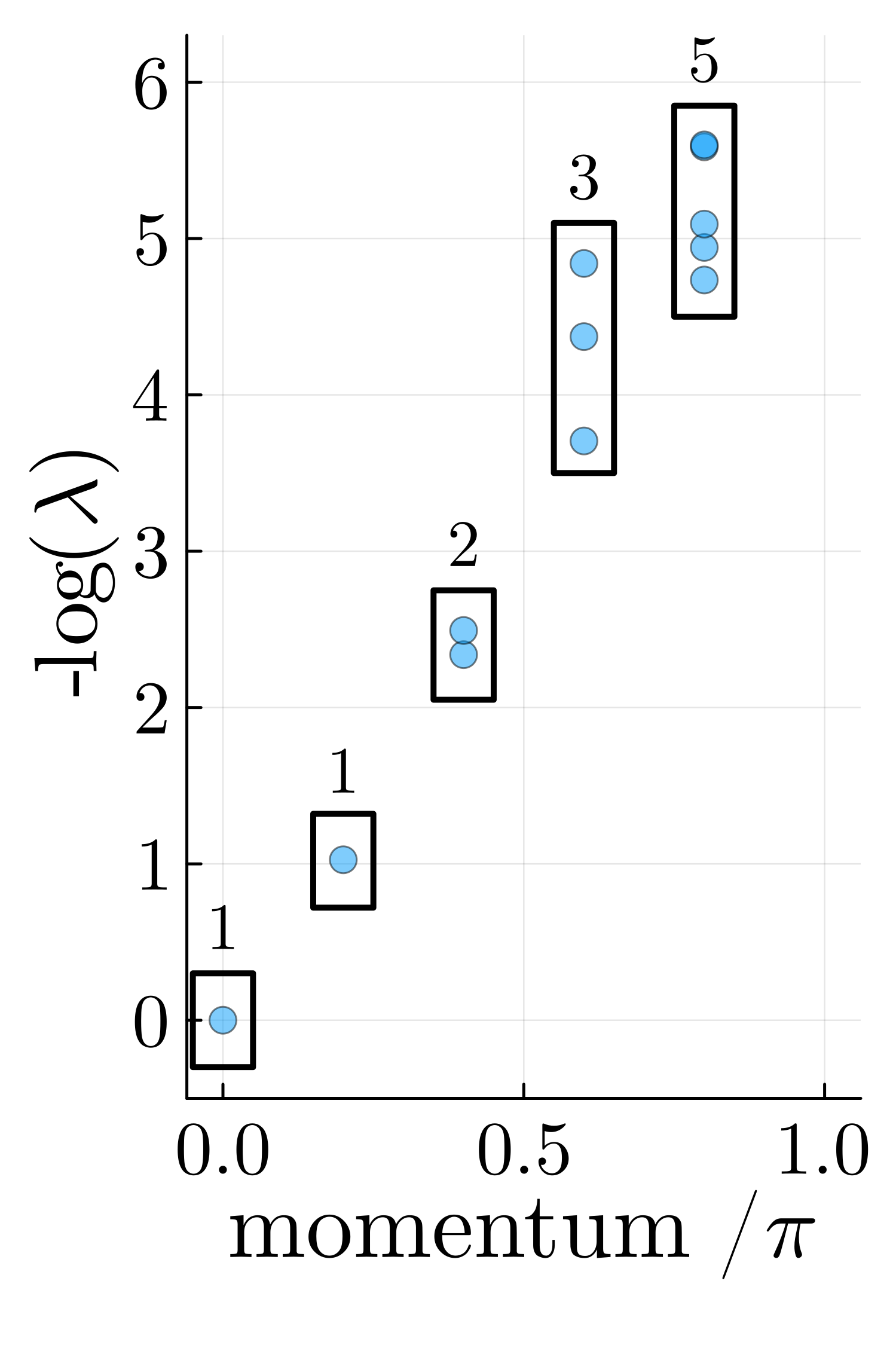}
        \caption{}
        \label{fig:ES cylinder}
    \end{subfigure}
    \caption{\raggedright \small  (a): The orange dots show the low lying part off the entanglement spectrum of a state on the incompressible plateau at low filling in fig. \ref{fig:occ_hardcore}. We clearly find a chiral entanglement spectrum indicating that this is a fractional quantum Hall state. The blue dots show the low lying part off the entanglement spectrum of a state on the incompressible plateau at high filling with reversed chirality. (b): The low lying part of the entanglement spectrum is shown for three values of the approximation parameter $\vartheta$, illustrating that we can systematically increase the chirality of the entanglement spectrum. (c): lowest branch of the entanglement spectrum on a 10 site cylinder. The counting associated with the chiral boson on the edge of the Laughlin state is observed.}

    \label{fig:enter-label}
\end{figure}
We initially consider the $U \xrightarrow{} \infty$ limit of hardcore bosons with a flux of $\phi = \pi/2$ per plaquette. 
Note, that for this choice of flux, the non-interacting Hamiltonian has remarkably flat lowest and highest bands, making it a promising choice for exhibiting fractional quantum Hall physics. 
The flatness of a band can be quantified by the ratio of its gap to the next band and its bandwidth \cite{Wen2011, Neupert2011, Sun2011}. 
Further, this large choice of flux is within the reach of current cold atom experiments \cite{Aidelsburger2011, Aidelsburger2013, Ketterle2013, Bloch2014chrialcurrent, Greiner2017microscopy, Greiner2023FQHrealization, aidelsburger2018review}.
To search for fractional Hall states in this model we first map out the average occupation per site $\langle \hat{n}\rangle$ of the ground state as a function of the chemical potential $\mu$, see fig. \ref{fig:occ_hardcore}. 
The calculations were performed with bulk bond dimensions $\chi_B = 4$ and $\chi_B = 5$ and values of environment bond dimension up to $\chi_E = 120$, which showed converged behaviour for the observables, as shown in figure \ref{fig:occ_hardcore}. 
We used the Landau gauge for the vector potential in eq.~\eqref{eq: Hamiltonian} and multiples of the $4\times1$ magnetic unit cell for the unit cell of the iPEPS ansatz, cf. supplementary material for more details on gauge choices and unit cells used in this letter \cite{SupplementaryMaterial}. 
We find two incompressible plateaus in the occupation, one at low and one at high filling. The states on these plateaus show a homogeneous density. The average occupation of the lower plateau corresponds exactly to the occupation for a $\frac{\langle \hat{n} \rangle}{2\pi \phi} = \nu = 1/2$ Laughlin state of bosons. 
We observe a symmetry of occupation around $\mu = 0$, cf fig. \ref{fig:occ_hardcore}. This can be understood by relating the Hamiltonian from eq.~\eqref{eq: Hamiltonian} for a particular choice of flux and chemical potential $\hat{H}(\phi, \mu)$ to the same Hamiltonian with reversed sign for the chemical potential $\hat{H}(\phi, -\mu)$ using a particle-hole transformation
\begin{equation}
    O_{\text{ph}} = \prod_i (a_i^{\dagger} + a_i)
\end{equation}
as well as time reversal $\Theta$. Applying these transformations results in
\begin{equation}
    \Theta O_{\text{ph}} \hat{H}(\phi, \mu) O_{\text{ph}} \Theta^{-1} = \hat{H}(\phi, -\mu) - \mu N,
\end{equation}\\
with $N$ the number of sites in the system. This also relates the eigenstates of $\hat{H}(\phi, \mu)$ and $\hat{H}(\phi, -\mu)$ and hence can be used to understand precise nature of the plateau at high filling of fig. \ref{fig:occ_hardcore} as a Laughlin state of holes at hole-filling $\Bar{\nu} = \frac{\langle \hat{n}_\text{h} \rangle}{2\pi \phi} = 1/2$, with $\langle \hat{n}_\text{h} \rangle = 1 - \langle \hat{n} \rangle$. We note further, that the incompressible plateaus in fig. \ref{fig:occ_hardcore} make up more than 20\% of the parameter-space between the empty and filled states. We attribute the fact that these plateaus make up such a sizable fraction to the flatness of the lower and upper bands of the Hamiltonian at $\phi = \pi / 2$. 

We now characterize the properties of the states at the $\nu = 1/2$ filling in order to verify that this is indeed a chiral Laughlin state.
We do so by first inspecting the entanglement spectrum, which should correspond to the gapless spectrum of the chiral edge modes, by following  Li and Haldane \cite{Li&Haldane2008}.
The entanglement spectrum is defined as the spectrum of the entanglement Hamiltonian $H_{ent}$ that, in turn, can identified from the reduced density matrix $\rho_l$ of a bipartition of the system $\rho_l = e^{-H_{ent}}$.
We can access the entanglement spectrum for our system by utilizing the bulk-boundary correspondence for PEPS \cite{Cirac2011BBC} and do so directly in the thermodynamic limit \cite{Haegeman2017Medley}  as well as on a cylinder of finite circumference. 
A more detailed description the methods used for the computation and details on the approximations made can be found in the supplementary material \cite{SupplementaryMaterial}.

In fig. \ref{fig:ES_combi} we can see that we find a almost perfectly chiral entanglement spectrum for the states on the incompressible plateau at the $\nu = 1/2$ filling, confirming the chiral nature of the edge modes of the states on the incompressible plateaus in fig. \ref{fig:occ_hardcore}.
For the the states on the incompressible plateau at high filling, which we expect to be a Laughlin state of holes, we find that they have a chiral entanglement spectrum with reversed chirality, see fig. \ref{fig:ES_combi}, as expected for holes. 
The fact that the entanglement spectra shown here are not perfectly chiral at momentum zero stems from the numerical approximations made, but the chirality can systematically be improved by increasing the approximation parameter $\vartheta$ as illustrated in fig. \ref{fig:ES parameters}. This parameter corresponds to the bond dimension of the boundary Hamiltonian from the PEPS bulk-boundary correspondence, cf. suppl. material for more information. We also perform the entanglement spectrum on a finite cylinder of circumference 10. This discretizes the edge spectrum and allows us to identify the degeneracy of the lowest energy edge excitations, cf. fig. \ref{fig:ES cylinder}. These degeneracies (1, 1, 2, 3, 5, ...) are the ones expected for a chiral boson modes, the edge mode of the Laughlin state.

\begin{figure}[h]
    \centering
    \includegraphics[width=0.4\textwidth]{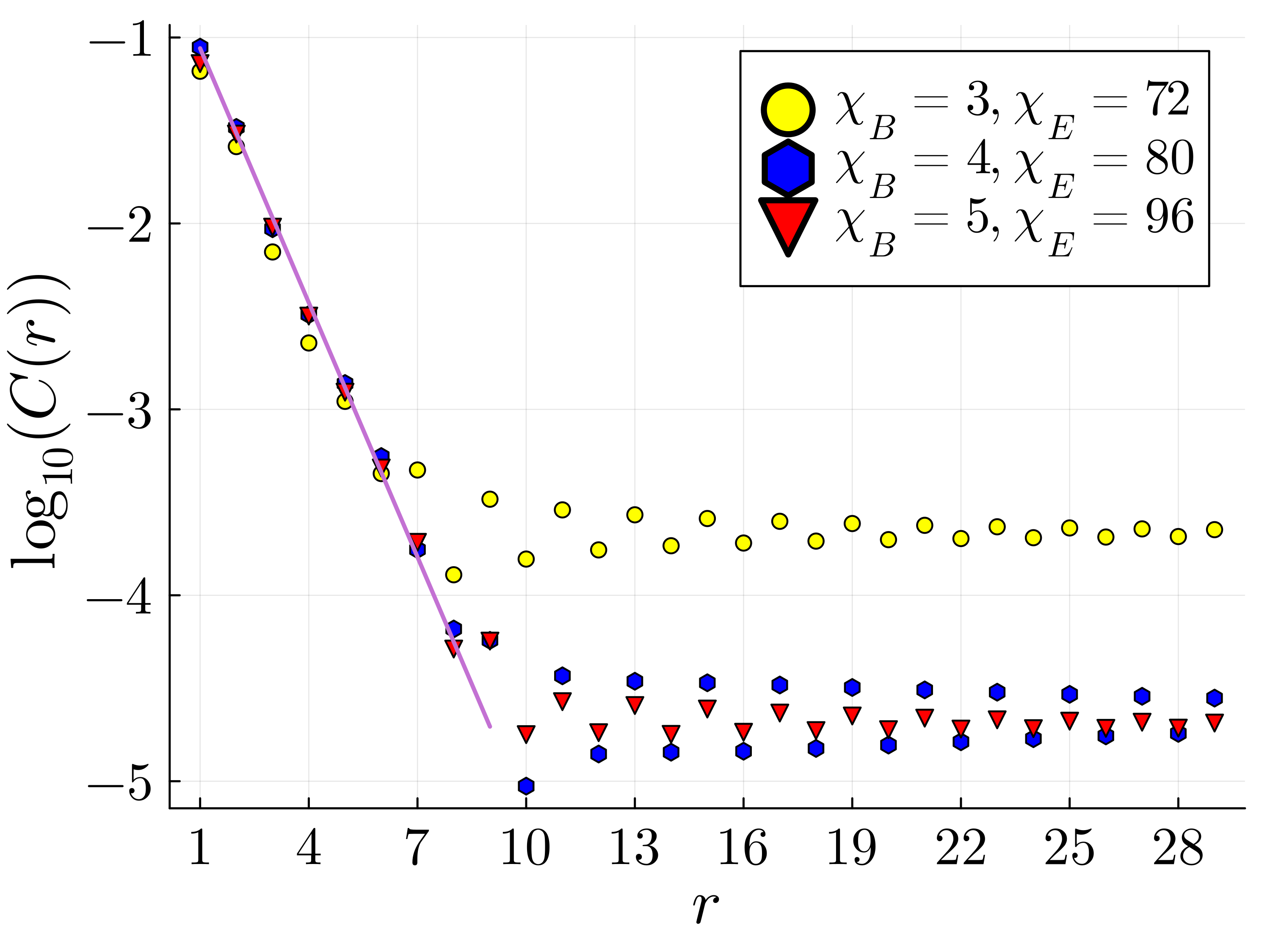}
    \caption{\raggedright \small Correlation function of the plateau state at low filling, cf. fig. \ref{fig:occ_hardcore}. At short distances the correlations decay exponentially with distance, consistent with a bulk gap of the state. We extract a correlation length of $\xi/\ell_B = 1.32$ which is consistent with previous results \cite{Rizzi2017TTN-FQH}. At long distances we find a small aritficial tail of in the correlation functions, that is due to the fact that we represent a chiral topological gapped state with PEPS.}
    \label{fig:correlation function}
\end{figure}

Additionally we also calculate bulk correlation functions for the incompressible states at $\nu = 1 /2$. We find a exponential decay of correlations at short distances, consistent with the bulk gap of the Laughlin states, with a correlation length of $\xi / \ell_B = 1.32$ in units of the magnetic length $\ell_B = 1 / \sqrt{\phi}$. This is compatible with the result obtained with tree tensor networks for finite systems at lower flux \cite{Rizzi2017TTN-FQH}. 
Furthermore, we observe a finite tail in the correlation function at long distances. This tail is a numerical artifact and can be identified as consequence of the fact that we are representing a gapped chiral state with PEPS: it has been observed as well in the context of related spin systems \cite{Hasik2022CSL}. Increasing the environment bond dimension $\chi_E$, for a fixed value of $\chi_B$ changes the details of the artificial long range tail, while leaving the short range exponentially decaying correlations invariant. The value at which these tails appear depends on $\chi_B$.

\begin{figure}[t]
    \centering
    \includegraphics[width=0.4\textwidth]{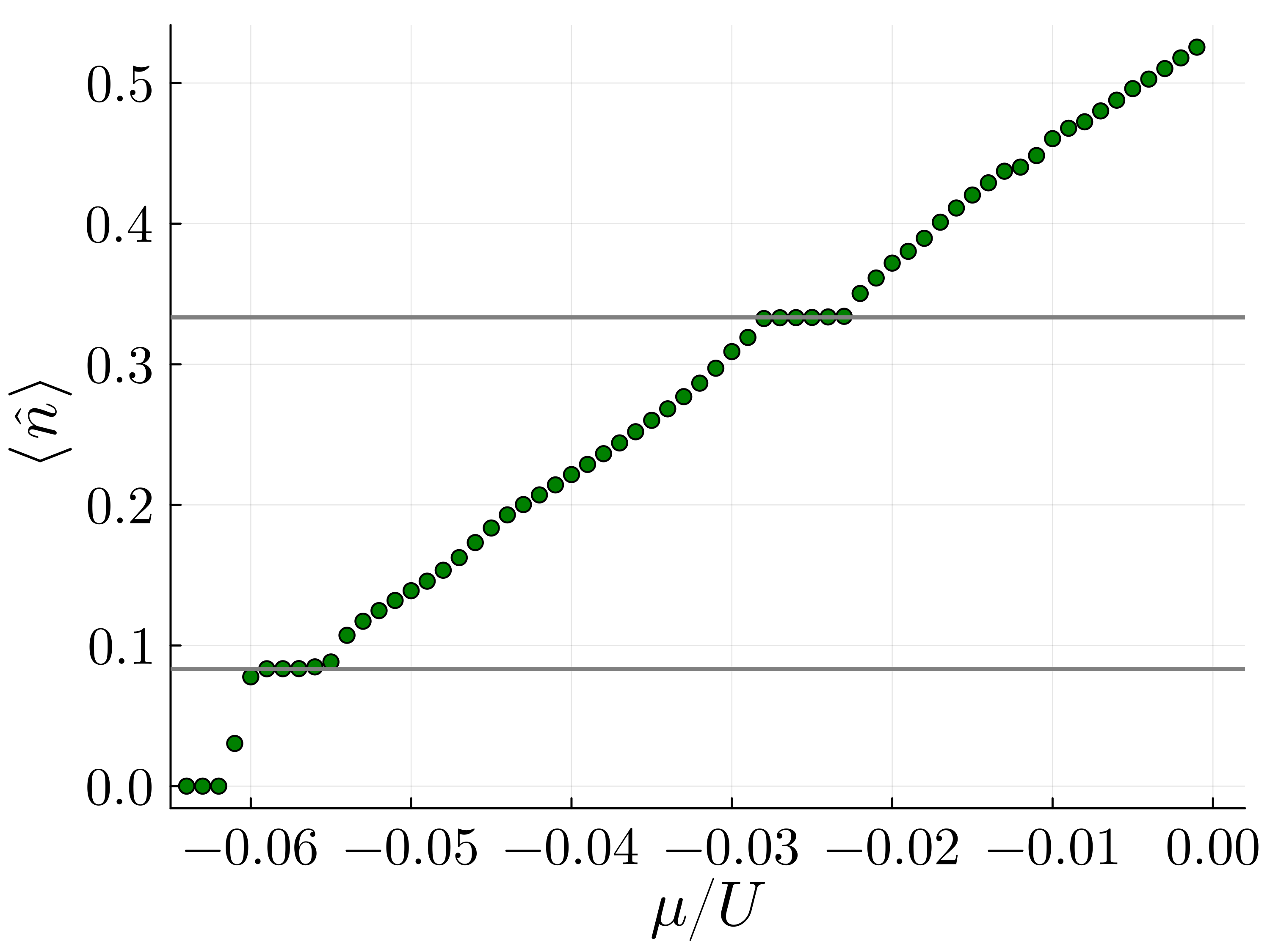}
    \caption{\raggedright \small Softcore boson occupation per site for $\frac{U}{t} = 50$ and flux $\phi = \pi / 3$. We see that an additional incompressible plateau at $\nu =2$. The plateau at $\nu = 2$ corresponds to bosonic integer quantum Hall states.}
    \label{fig:occ_softcore1}
\end{figure}

In order to show that we are not limited to hard-core bosons we
relax the $U \xrightarrow{} \infty$ condition and move to large but finite interactions $U/t = 50$.  In our calculations we allow up to two bosons per site in this case, given that even higher fillings are energetically strongly suppressed.
In this situation we observe, for a magnetic flux of \mbox{$\phi = \pi/2$}, a similar average occupation per site as in the case of hardcore bosons, with a slightly higher occupation at a given $\mu$ stemming from the fact that states including two bosons on a single site are now possible. 
We again find incompressible plateaus at $\nu = 1/2$ as well as at its hole analog, cf. supplementary material \cite{SupplementaryMaterial}. 
If we instead reduce the magnetic flux to $\phi = \pi/3$, we additionally find a plateau at filling factor $\nu =2$. 
This plateau is expected to be associated with exotic states that belong to the \textit{Jain sequence} obtained in the composite fermion picture \cite{Jain1989}.
The $\nu = 2$ state from this sequence is a \textit{bosonic integer quantum hall} state, which is not intrinsically topological but rather a symmetry protected topological state \cite{Senthil2013BIQH, Furukawa2013BIQH, He2015BIQH, Moller2015, Sterdyniak2015BIQH}.


\textit{Conclustion \& Outlook.}
$-$
In this letter we demonstrated for the first time that the variational iPEPS framework allows for the investigation of Hamiltonians of mobile, interacting particles with semi-classical gauge fields that host chiral topological states as ground states, such as fractional Hall states and Chern insulators. 
The iPEPS method is a crucial complement to other well established methods for this task (such as exact diagonalization or other tensor network ansatzes) as it avoids finite size effects that can alter the correct phase diagram.
Further, as it can give genuine information about the bulk physics, it can be used to benchmark whether finite-size experiments are actually displaying bulk behaviour or are dominated by their boundary.
The class of Hamiltonians describing mobile, interacting particles in a semi-classical gauge field, of which we have treated a paradigmatic example in this letter, comes up in the description of several physical systems that have attracted enormous amounts of interest in recent years \cite{Aidelsburger2013, Greiner2023FQHrealization, ponomarenko2013cloning, bergholtz2013FCI, Vishvanath2021, Wu2023}.
Furthermore key characteristics, such as the importance of flat bands which aid the emergence of interesting correlated phases, are often shared among these different systems.
In the bosonic Harper-Hofstadter model, the (relatively) flat bands of the Hamiltonian can be understood as leading to the prominence of interaction driven physics and hence to fractional Hall states~\cite{Parameswaran2013fractional}.
Correspondingly, in twisted bilayers systems the isolated moiré flat bands are the reason for many candidate ground states such as spin-liquid states, quantum anomalous Hall insulators, and chiral d-wave superconductors \cite{MacDonald2018}.
An important illustrative example is twisted $\ce{WSe2}$ which can be described by a moiré Hubbard model \cite{Wang2020,kiese2022tmds}, in which the hopping on the triangular lattice is enhanced by a spin-dependent phase, tunable in experiments. 
Given the similarity of this Hamiltonian to the one studied in this letter, this puts the investigation of the physics of twisted bilayer systems well within reach of the variational iPEPS framework, promising an unbiased numerical perspective on the physics of these systems.
Furthermore, recent advances drastically improved the ability to calculate excitation spectra  and structure factors in the variational iPEPS framework \cite{Ponsioen22spectalfunctions, ponsioen2023improvedSF, tu2023generating}. This allows direct connection to experimental results \cite{Heisenberg22spectrum} and thus should establish the variational iPEPS framework clearly as a mayor tool of investigation into the above mentioned systems.

\textit{Acknowledgements.}
$-$
We thank Philipp Schmoll, Niklas Tausendpfund and Simon Trebst for discussions and the group of Jens Eisert for the warm hospitality during the visit supported by the CRC 183.
E.\ L.\ W.\ thanks the Studienstiftung des deutschen Volkes for support.
The authors would like to warmly endorse the open source Tensor Network libraries of Quantum-Ghent group and acknowledge in particular the use of \textsc{MPSKit} for the entanglement spectrum calculations.
This work has been funded by the Deutsche
Forschungsgemeinschaft (DFG, German Research Foundation) under the project number 277101999 – CRC 183 (project B01).
The authors gratefully acknowledge the Gauss Centre for Supercomputing e.V. (www.gauss-centre.eu) for funding this project by providing computing time through the John von Neumann Institute for Computing (NIC) on the GCS Supercomputer JUWELS at the J\"ulich Supercomputing Centre (JSC) (Grant NeTeNeSyQuMa) and the FZ J\"ulich for JURECA (institute project PGI-8)~\cite{JURECA2021}.\\


\bibliography{references}

\begin{thebibliography}{79}%
\makeatletter
\providecommand \@ifxundefined [1]{%
 \@ifx{#1\undefined}
}%
\providecommand \@ifnum [1]{%
 \ifnum #1\expandafter \@firstoftwo
 \else \expandafter \@secondoftwo
 \fi
}%
\providecommand \@ifx [1]{%
 \ifx #1\expandafter \@firstoftwo
 \else \expandafter \@secondoftwo
 \fi
}%
\providecommand \natexlab [1]{#1}%
\providecommand \enquote  [1]{``#1''}%
\providecommand \bibnamefont  [1]{#1}%
\providecommand \bibfnamefont [1]{#1}%
\providecommand \citenamefont [1]{#1}%
\providecommand \href@noop [0]{\@secondoftwo}%
\providecommand \href [0]{\begingroup \@sanitize@url \@href}%
\providecommand \@href[1]{\@@startlink{#1}\@@href}%
\providecommand \@@href[1]{\endgroup#1\@@endlink}%
\providecommand \@sanitize@url [0]{\catcode `\\12\catcode `\$12\catcode `\&12\catcode `\#12\catcode `\^12\catcode `\_12\catcode `\%12\relax}%
\providecommand \@@startlink[1]{}%
\providecommand \@@endlink[0]{}%
\providecommand \url  [0]{\begingroup\@sanitize@url \@url }%
\providecommand \@url [1]{\endgroup\@href {#1}{\urlprefix }}%
\providecommand \urlprefix  [0]{URL }%
\providecommand \Eprint [0]{\href }%
\providecommand \doibase [0]{https://doi.org/}%
\providecommand \selectlanguage [0]{\@gobble}%
\providecommand \bibinfo  [0]{\@secondoftwo}%
\providecommand \bibfield  [0]{\@secondoftwo}%
\providecommand \translation [1]{[#1]}%
\providecommand \BibitemOpen [0]{}%
\providecommand \bibitemStop [0]{}%
\providecommand \bibitemNoStop [0]{.\EOS\space}%
\providecommand \EOS [0]{\spacefactor3000\relax}%
\providecommand \BibitemShut  [1]{\csname bibitem#1\endcsname}%
\let\auto@bib@innerbib\@empty
\bibitem [{\citenamefont {Wen}(2017)}]{Wen2017ZOO}%
  \BibitemOpen
  \bibfield  {author} {\bibinfo {author} {\bibfnamefont {X.-G.}\ \bibnamefont {Wen}},\ }\bibfield  {title} {\bibinfo {title} {Colloquium: Zoo of quantum-topological phases of matter},\ }\href {https://doi.org/10.1103/RevModPhys.89.041004} {\bibfield  {journal} {\bibinfo  {journal} {Rev. Mod. Phys.}\ }\textbf {\bibinfo {volume} {89}},\ \bibinfo {pages} {041004} (\bibinfo {year} {2017})}\BibitemShut {NoStop}%
\bibitem [{\citenamefont {Kitaev}(2003)}]{kitaev2003fault}%
  \BibitemOpen
  \bibfield  {author} {\bibinfo {author} {\bibfnamefont {A.~Y.}\ \bibnamefont {Kitaev}},\ }\bibfield  {title} {\bibinfo {title} {Fault-tolerant quantum computation by anyons},\ }\href@noop {} {\bibfield  {journal} {\bibinfo  {journal} {Annals of physics}\ }\textbf {\bibinfo {volume} {303}},\ \bibinfo {pages} {2} (\bibinfo {year} {2003})}\BibitemShut {NoStop}%
\bibitem [{\citenamefont {Nayak}\ \emph {et~al.}(2008)\citenamefont {Nayak}, \citenamefont {Simon}, \citenamefont {Stern}, \citenamefont {Freedman},\ and\ \citenamefont {Das~Sarma}}]{DasSarma2008TQC}%
  \BibitemOpen
  \bibfield  {author} {\bibinfo {author} {\bibfnamefont {C.}~\bibnamefont {Nayak}}, \bibinfo {author} {\bibfnamefont {S.~H.}\ \bibnamefont {Simon}}, \bibinfo {author} {\bibfnamefont {A.}~\bibnamefont {Stern}}, \bibinfo {author} {\bibfnamefont {M.}~\bibnamefont {Freedman}},\ and\ \bibinfo {author} {\bibfnamefont {S.}~\bibnamefont {Das~Sarma}},\ }\bibfield  {title} {\bibinfo {title} {Non-abelian anyons and topological quantum computation},\ }\href {https://doi.org/10.1103/RevModPhys.80.1083} {\bibfield  {journal} {\bibinfo  {journal} {Rev. Mod. Phys.}\ }\textbf {\bibinfo {volume} {80}},\ \bibinfo {pages} {1083} (\bibinfo {year} {2008})}\BibitemShut {NoStop}%
\bibitem [{\citenamefont {Tsui}\ \emph {et~al.}(1982)\citenamefont {Tsui}, \citenamefont {Stormer},\ and\ \citenamefont {Gossard}}]{Stromer82FQH}%
  \BibitemOpen
  \bibfield  {author} {\bibinfo {author} {\bibfnamefont {D.~C.}\ \bibnamefont {Tsui}}, \bibinfo {author} {\bibfnamefont {H.~L.}\ \bibnamefont {Stormer}},\ and\ \bibinfo {author} {\bibfnamefont {A.~C.}\ \bibnamefont {Gossard}},\ }\bibfield  {title} {\bibinfo {title} {Two-dimensional magnetotransport in the extreme quantum limit},\ }\href {https://doi.org/10.1103/PhysRevLett.48.1559} {\bibfield  {journal} {\bibinfo  {journal} {Phys. Rev. Lett.}\ }\textbf {\bibinfo {volume} {48}},\ \bibinfo {pages} {1559} (\bibinfo {year} {1982})}\BibitemShut {NoStop}%
\bibitem [{\citenamefont {Laughlin}(1983)}]{Laughlin83FQH}%
  \BibitemOpen
  \bibfield  {author} {\bibinfo {author} {\bibfnamefont {R.~B.}\ \bibnamefont {Laughlin}},\ }\bibfield  {title} {\bibinfo {title} {Anomalous quantum hall effect: An incompressible quantum fluid with fractionally charged excitations},\ }\href {https://doi.org/10.1103/PhysRevLett.50.1395} {\bibfield  {journal} {\bibinfo  {journal} {Phys. Rev. Lett.}\ }\textbf {\bibinfo {volume} {50}},\ \bibinfo {pages} {1395} (\bibinfo {year} {1983})}\BibitemShut {NoStop}%
\bibitem [{\citenamefont {Parameswaran}\ \emph {et~al.}(2013{\natexlab{a}})\citenamefont {Parameswaran}, \citenamefont {Roy},\ and\ \citenamefont {Sondhi}}]{Parameswaran2013}%
  \BibitemOpen
  \bibfield  {author} {\bibinfo {author} {\bibfnamefont {S.~A.}\ \bibnamefont {Parameswaran}}, \bibinfo {author} {\bibfnamefont {R.}~\bibnamefont {Roy}},\ and\ \bibinfo {author} {\bibfnamefont {S.~L.}\ \bibnamefont {Sondhi}},\ }\bibfield  {title} {\bibinfo {title} {Fractional quantum hall physics in topological flat bands},\ }\href {https://doi.org/https://doi.org/10.1016/j.crhy.2013.04.003} {\bibfield  {journal} {\bibinfo  {journal} {Comptes Rendus Physique}\ }\textbf {\bibinfo {volume} {14}},\ \bibinfo {pages} {816} (\bibinfo {year} {2013}{\natexlab{a}})},\ \bibinfo {note} {topological insulators / Isolants topologiques}\BibitemShut {NoStop}%
\bibitem [{\citenamefont {Aidelsburger}\ \emph {et~al.}(2011)\citenamefont {Aidelsburger}, \citenamefont {Atala}, \citenamefont {Nascimb\`ene}, \citenamefont {Trotzky}, \citenamefont {Chen},\ and\ \citenamefont {Bloch}}]{Aidelsburger2011}%
  \BibitemOpen
  \bibfield  {author} {\bibinfo {author} {\bibfnamefont {M.}~\bibnamefont {Aidelsburger}}, \bibinfo {author} {\bibfnamefont {M.}~\bibnamefont {Atala}}, \bibinfo {author} {\bibfnamefont {S.}~\bibnamefont {Nascimb\`ene}}, \bibinfo {author} {\bibfnamefont {S.}~\bibnamefont {Trotzky}}, \bibinfo {author} {\bibfnamefont {Y.-A.}\ \bibnamefont {Chen}},\ and\ \bibinfo {author} {\bibfnamefont {I.}~\bibnamefont {Bloch}},\ }\bibfield  {title} {\bibinfo {title} {Experimental realization of strong effective magnetic fields in an optical lattice},\ }\href {https://doi.org/10.1103/PhysRevLett.107.255301} {\bibfield  {journal} {\bibinfo  {journal} {Phys. Rev. Lett.}\ }\textbf {\bibinfo {volume} {107}},\ \bibinfo {pages} {255301} (\bibinfo {year} {2011})}\BibitemShut {NoStop}%
\bibitem [{\citenamefont {Aidelsburger}\ \emph {et~al.}(2013)\citenamefont {Aidelsburger}, \citenamefont {Atala}, \citenamefont {Lohse}, \citenamefont {Barreiro}, \citenamefont {Paredes},\ and\ \citenamefont {Bloch}}]{Aidelsburger2013}%
  \BibitemOpen
  \bibfield  {author} {\bibinfo {author} {\bibfnamefont {M.}~\bibnamefont {Aidelsburger}}, \bibinfo {author} {\bibfnamefont {M.}~\bibnamefont {Atala}}, \bibinfo {author} {\bibfnamefont {M.}~\bibnamefont {Lohse}}, \bibinfo {author} {\bibfnamefont {J.~T.}\ \bibnamefont {Barreiro}}, \bibinfo {author} {\bibfnamefont {B.}~\bibnamefont {Paredes}},\ and\ \bibinfo {author} {\bibfnamefont {I.}~\bibnamefont {Bloch}},\ }\bibfield  {title} {\bibinfo {title} {Realization of the hofstadter hamiltonian with ultracold atoms in optical lattices},\ }\href {https://doi.org/10.1103/PhysRevLett.111.185301} {\bibfield  {journal} {\bibinfo  {journal} {Phys. Rev. Lett.}\ }\textbf {\bibinfo {volume} {111}},\ \bibinfo {pages} {185301} (\bibinfo {year} {2013})}\BibitemShut {NoStop}%
\bibitem [{\citenamefont {Miyake}\ \emph {et~al.}(2013)\citenamefont {Miyake}, \citenamefont {Siviloglou}, \citenamefont {Kennedy}, \citenamefont {Burton},\ and\ \citenamefont {Ketterle}}]{Ketterle2013}%
  \BibitemOpen
  \bibfield  {author} {\bibinfo {author} {\bibfnamefont {H.}~\bibnamefont {Miyake}}, \bibinfo {author} {\bibfnamefont {G.~A.}\ \bibnamefont {Siviloglou}}, \bibinfo {author} {\bibfnamefont {C.~J.}\ \bibnamefont {Kennedy}}, \bibinfo {author} {\bibfnamefont {W.~C.}\ \bibnamefont {Burton}},\ and\ \bibinfo {author} {\bibfnamefont {W.}~\bibnamefont {Ketterle}},\ }\bibfield  {title} {\bibinfo {title} {Realizing the harper hamiltonian with laser-assisted tunneling in optical lattices},\ }\href {https://doi.org/10.1103/PhysRevLett.111.185302} {\bibfield  {journal} {\bibinfo  {journal} {Phys. Rev. Lett.}\ }\textbf {\bibinfo {volume} {111}},\ \bibinfo {pages} {185302} (\bibinfo {year} {2013})}\BibitemShut {NoStop}%
\bibitem [{\citenamefont {Atala}\ \emph {et~al.}(2014)\citenamefont {Atala}, \citenamefont {Aidelsburger}, \citenamefont {Lohse}, \citenamefont {Barreiro}, \citenamefont {Paredes},\ and\ \citenamefont {Bloch}}]{Bloch2014chrialcurrent}%
  \BibitemOpen
  \bibfield  {author} {\bibinfo {author} {\bibfnamefont {M.}~\bibnamefont {Atala}}, \bibinfo {author} {\bibfnamefont {M.}~\bibnamefont {Aidelsburger}}, \bibinfo {author} {\bibfnamefont {M.}~\bibnamefont {Lohse}}, \bibinfo {author} {\bibfnamefont {J.~T.}\ \bibnamefont {Barreiro}}, \bibinfo {author} {\bibfnamefont {B.}~\bibnamefont {Paredes}},\ and\ \bibinfo {author} {\bibfnamefont {I.}~\bibnamefont {Bloch}},\ }\bibfield  {title} {\bibinfo {title} {Observation of chiral currents with ultracold atoms in bosonic ladders},\ }\href@noop {} {\bibfield  {journal} {\bibinfo  {journal} {Nature Physics}\ }\textbf {\bibinfo {volume} {10}},\ \bibinfo {pages} {588} (\bibinfo {year} {2014})}\BibitemShut {NoStop}%
\bibitem [{\citenamefont {Tai}\ \emph {et~al.}(2017)\citenamefont {Tai}, \citenamefont {Lukin}, \citenamefont {Rispoli}, \citenamefont {Schittko}, \citenamefont {Menke}, \citenamefont {Borgnia}, \citenamefont {Preiss}, \citenamefont {Grusdt}, \citenamefont {Kaufman},\ and\ \citenamefont {Greiner}}]{Greiner2017microscopy}%
  \BibitemOpen
  \bibfield  {author} {\bibinfo {author} {\bibfnamefont {M.~E.}\ \bibnamefont {Tai}}, \bibinfo {author} {\bibfnamefont {A.}~\bibnamefont {Lukin}}, \bibinfo {author} {\bibfnamefont {M.}~\bibnamefont {Rispoli}}, \bibinfo {author} {\bibfnamefont {R.}~\bibnamefont {Schittko}}, \bibinfo {author} {\bibfnamefont {T.}~\bibnamefont {Menke}}, \bibinfo {author} {\bibfnamefont {D.}~\bibnamefont {Borgnia}}, \bibinfo {author} {\bibfnamefont {P.~M.}\ \bibnamefont {Preiss}}, \bibinfo {author} {\bibfnamefont {F.}~\bibnamefont {Grusdt}}, \bibinfo {author} {\bibfnamefont {A.~M.}\ \bibnamefont {Kaufman}},\ and\ \bibinfo {author} {\bibfnamefont {M.}~\bibnamefont {Greiner}},\ }\bibfield  {title} {\bibinfo {title} {Microscopy of the interacting harper--hofstadter model in the two-body limit},\ }\href@noop {} {\bibfield  {journal} {\bibinfo  {journal} {Nature}\ }\textbf {\bibinfo {volume} {546}},\ \bibinfo {pages} {519} (\bibinfo {year} {2017})}\BibitemShut {NoStop}%
\bibitem [{\citenamefont {L{\'e}onard}\ \emph {et~al.}(2023)\citenamefont {L{\'e}onard}, \citenamefont {Kim}, \citenamefont {Kwan}, \citenamefont {Segura}, \citenamefont {Grusdt}, \citenamefont {Repellin}, \citenamefont {Goldman},\ and\ \citenamefont {Greiner}}]{Greiner2023FQHrealization}%
  \BibitemOpen
  \bibfield  {author} {\bibinfo {author} {\bibfnamefont {J.}~\bibnamefont {L{\'e}onard}}, \bibinfo {author} {\bibfnamefont {S.}~\bibnamefont {Kim}}, \bibinfo {author} {\bibfnamefont {J.}~\bibnamefont {Kwan}}, \bibinfo {author} {\bibfnamefont {P.}~\bibnamefont {Segura}}, \bibinfo {author} {\bibfnamefont {F.}~\bibnamefont {Grusdt}}, \bibinfo {author} {\bibfnamefont {C.}~\bibnamefont {Repellin}}, \bibinfo {author} {\bibfnamefont {N.}~\bibnamefont {Goldman}},\ and\ \bibinfo {author} {\bibfnamefont {M.}~\bibnamefont {Greiner}},\ }\bibfield  {title} {\bibinfo {title} {Realization of a fractional quantum hall state with ultracold atoms},\ }\href@noop {} {\bibfield  {journal} {\bibinfo  {journal} {Nature}\ ,\ \bibinfo {pages} {1}} (\bibinfo {year} {2023})}\BibitemShut {NoStop}%
\bibitem [{\citenamefont {Verstraete}\ and\ \citenamefont {Cirac}(2004)}]{verstraete2004PEPS}%
  \BibitemOpen
  \bibfield  {author} {\bibinfo {author} {\bibfnamefont {F.}~\bibnamefont {Verstraete}}\ and\ \bibinfo {author} {\bibfnamefont {J.~I.}\ \bibnamefont {Cirac}},\ }\href@noop {} {\bibinfo {title} {Renormalization algorithms for quantum-many body systems in two and higher dimensions}} (\bibinfo {year} {2004}),\ \Eprint {https://arxiv.org/abs/cond-mat/0407066} {arXiv:cond-mat/0407066 [cond-mat.str-el]} \BibitemShut {NoStop}%
\bibitem [{\citenamefont {Dubail}\ and\ \citenamefont {Read}(2015)}]{Read2015NOGO}%
  \BibitemOpen
  \bibfield  {author} {\bibinfo {author} {\bibfnamefont {J.}~\bibnamefont {Dubail}}\ and\ \bibinfo {author} {\bibfnamefont {N.}~\bibnamefont {Read}},\ }\bibfield  {title} {\bibinfo {title} {Tensor network trial states for chiral topological phases in two dimensions and a no-go theorem in any dimension},\ }\href {https://doi.org/10.1103/PhysRevB.92.205307} {\bibfield  {journal} {\bibinfo  {journal} {Phys. Rev. B}\ }\textbf {\bibinfo {volume} {92}},\ \bibinfo {pages} {205307} (\bibinfo {year} {2015})}\BibitemShut {NoStop}%
\bibitem [{\citenamefont {Hasik}\ \emph {et~al.}(2022{\natexlab{a}})\citenamefont {Hasik}, \citenamefont {Van~Damme}, \citenamefont {Poilblanc},\ and\ \citenamefont {Vanderstraeten}}]{Hasik2022CSL}%
  \BibitemOpen
  \bibfield  {author} {\bibinfo {author} {\bibfnamefont {J.}~\bibnamefont {Hasik}}, \bibinfo {author} {\bibfnamefont {M.}~\bibnamefont {Van~Damme}}, \bibinfo {author} {\bibfnamefont {D.}~\bibnamefont {Poilblanc}},\ and\ \bibinfo {author} {\bibfnamefont {L.}~\bibnamefont {Vanderstraeten}},\ }\bibfield  {title} {\bibinfo {title} {Simulating chiral spin liquids with projected entangled-pair states},\ }\href {https://doi.org/10.1103/PhysRevLett.129.177201} {\bibfield  {journal} {\bibinfo  {journal} {Phys. Rev. Lett.}\ }\textbf {\bibinfo {volume} {129}},\ \bibinfo {pages} {177201} (\bibinfo {year} {2022}{\natexlab{a}})}\BibitemShut {NoStop}%
\bibitem [{\citenamefont {Nielsen}\ \emph {et~al.}(2012)\citenamefont {Nielsen}, \citenamefont {Cirac},\ and\ \citenamefont {Sierra}}]{Nielson2012CSLparent1}%
  \BibitemOpen
  \bibfield  {author} {\bibinfo {author} {\bibfnamefont {A.~E.~B.}\ \bibnamefont {Nielsen}}, \bibinfo {author} {\bibfnamefont {J.~I.}\ \bibnamefont {Cirac}},\ and\ \bibinfo {author} {\bibfnamefont {G.}~\bibnamefont {Sierra}},\ }\bibfield  {title} {\bibinfo {title} {Laughlin spin-liquid states on lattices obtained from conformal field theory},\ }\href {https://doi.org/10.1103/PhysRevLett.108.257206} {\bibfield  {journal} {\bibinfo  {journal} {Phys. Rev. Lett.}\ }\textbf {\bibinfo {volume} {108}},\ \bibinfo {pages} {257206} (\bibinfo {year} {2012})}\BibitemShut {NoStop}%
\bibitem [{\citenamefont {Nielsen}\ \emph {et~al.}(2013)\citenamefont {Nielsen}, \citenamefont {Sierra},\ and\ \citenamefont {Cirac}}]{Nielsen2013CSLparentlocal}%
  \BibitemOpen
  \bibfield  {author} {\bibinfo {author} {\bibfnamefont {A.~E.}\ \bibnamefont {Nielsen}}, \bibinfo {author} {\bibfnamefont {G.}~\bibnamefont {Sierra}},\ and\ \bibinfo {author} {\bibfnamefont {J.~I.}\ \bibnamefont {Cirac}},\ }\bibfield  {title} {\bibinfo {title} {Local models of fractional quantum hall states in lattices and physical implementation},\ }\href@noop {} {\bibfield  {journal} {\bibinfo  {journal} {Nature communications}\ }\textbf {\bibinfo {volume} {4}},\ \bibinfo {pages} {2864} (\bibinfo {year} {2013})}\BibitemShut {NoStop}%
\bibitem [{\citenamefont {Fazio}\ and\ \citenamefont {Van Der~Zant}(2001)}]{fazio2001JJA}%
  \BibitemOpen
  \bibfield  {author} {\bibinfo {author} {\bibfnamefont {R.}~\bibnamefont {Fazio}}\ and\ \bibinfo {author} {\bibfnamefont {H.}~\bibnamefont {Van Der~Zant}},\ }\bibfield  {title} {\bibinfo {title} {Quantum phase transitions and vortex dynamics in superconducting networks},\ }\href@noop {} {\bibfield  {journal} {\bibinfo  {journal} {Physics Reports}\ }\textbf {\bibinfo {volume} {355}},\ \bibinfo {pages} {235} (\bibinfo {year} {2001})}\BibitemShut {NoStop}%
\bibitem [{\citenamefont {Read}\ and\ \citenamefont {Rezayi}(1999)}]{Read1999}%
  \BibitemOpen
  \bibfield  {author} {\bibinfo {author} {\bibfnamefont {N.}~\bibnamefont {Read}}\ and\ \bibinfo {author} {\bibfnamefont {E.}~\bibnamefont {Rezayi}},\ }\bibfield  {title} {\bibinfo {title} {Beyond paired quantum hall states: Parafermions and incompressible states in the first excited landau level},\ }\href {https://doi.org/10.1103/PhysRevB.59.8084} {\bibfield  {journal} {\bibinfo  {journal} {Phys. Rev. B}\ }\textbf {\bibinfo {volume} {59}},\ \bibinfo {pages} {8084} (\bibinfo {year} {1999})}\BibitemShut {NoStop}%
\bibitem [{\citenamefont {S\o{}rensen}\ \emph {et~al.}(2005)\citenamefont {S\o{}rensen}, \citenamefont {Demler},\ and\ \citenamefont {Lukin}}]{Lukin2005}%
  \BibitemOpen
  \bibfield  {author} {\bibinfo {author} {\bibfnamefont {A.~S.}\ \bibnamefont {S\o{}rensen}}, \bibinfo {author} {\bibfnamefont {E.}~\bibnamefont {Demler}},\ and\ \bibinfo {author} {\bibfnamefont {M.~D.}\ \bibnamefont {Lukin}},\ }\bibfield  {title} {\bibinfo {title} {Fractional quantum hall states of atoms in optical lattices},\ }\href {https://doi.org/10.1103/PhysRevLett.94.086803} {\bibfield  {journal} {\bibinfo  {journal} {Phys. Rev. Lett.}\ }\textbf {\bibinfo {volume} {94}},\ \bibinfo {pages} {086803} (\bibinfo {year} {2005})}\BibitemShut {NoStop}%
\bibitem [{\citenamefont {Hafezi}\ \emph {et~al.}(2007)\citenamefont {Hafezi}, \citenamefont {S\o{}rensen}, \citenamefont {Demler},\ and\ \citenamefont {Lukin}}]{Lukin2007}%
  \BibitemOpen
  \bibfield  {author} {\bibinfo {author} {\bibfnamefont {M.}~\bibnamefont {Hafezi}}, \bibinfo {author} {\bibfnamefont {A.~S.}\ \bibnamefont {S\o{}rensen}}, \bibinfo {author} {\bibfnamefont {E.}~\bibnamefont {Demler}},\ and\ \bibinfo {author} {\bibfnamefont {M.~D.}\ \bibnamefont {Lukin}},\ }\bibfield  {title} {\bibinfo {title} {Fractional quantum hall effect in optical lattices},\ }\href {https://doi.org/10.1103/PhysRevA.76.023613} {\bibfield  {journal} {\bibinfo  {journal} {Phys. Rev. A}\ }\textbf {\bibinfo {volume} {76}},\ \bibinfo {pages} {023613} (\bibinfo {year} {2007})}\BibitemShut {NoStop}%
\bibitem [{\citenamefont {M\"oller}\ and\ \citenamefont {Cooper}(2009)}]{Moller2009BIQH}%
  \BibitemOpen
  \bibfield  {author} {\bibinfo {author} {\bibfnamefont {G.}~\bibnamefont {M\"oller}}\ and\ \bibinfo {author} {\bibfnamefont {N.~R.}\ \bibnamefont {Cooper}},\ }\bibfield  {title} {\bibinfo {title} {Composite fermion theory for bosonic quantum hall states on lattices},\ }\href {https://doi.org/10.1103/PhysRevLett.103.105303} {\bibfield  {journal} {\bibinfo  {journal} {Phys. Rev. Lett.}\ }\textbf {\bibinfo {volume} {103}},\ \bibinfo {pages} {105303} (\bibinfo {year} {2009})}\BibitemShut {NoStop}%
\bibitem [{\citenamefont {Mazza}\ \emph {et~al.}(2010)\citenamefont {Mazza}, \citenamefont {Rizzi}, \citenamefont {Lewenstein},\ and\ \citenamefont {Cirac}}]{Rizzi2010}%
  \BibitemOpen
  \bibfield  {author} {\bibinfo {author} {\bibfnamefont {L.}~\bibnamefont {Mazza}}, \bibinfo {author} {\bibfnamefont {M.}~\bibnamefont {Rizzi}}, \bibinfo {author} {\bibfnamefont {M.}~\bibnamefont {Lewenstein}},\ and\ \bibinfo {author} {\bibfnamefont {J.~I.}\ \bibnamefont {Cirac}},\ }\bibfield  {title} {\bibinfo {title} {Emerging bosons with three-body interactions from spin-1 atoms in optical lattices},\ }\href {https://doi.org/10.1103/PhysRevA.82.043629} {\bibfield  {journal} {\bibinfo  {journal} {Phys. Rev. A}\ }\textbf {\bibinfo {volume} {82}},\ \bibinfo {pages} {043629} (\bibinfo {year} {2010})}\BibitemShut {NoStop}%
\bibitem [{\citenamefont {Sterdyniak}\ \emph {et~al.}(2012)\citenamefont {Sterdyniak}, \citenamefont {Regnault},\ and\ \citenamefont {M\"oller}}]{Moller2012}%
  \BibitemOpen
  \bibfield  {author} {\bibinfo {author} {\bibfnamefont {A.}~\bibnamefont {Sterdyniak}}, \bibinfo {author} {\bibfnamefont {N.}~\bibnamefont {Regnault}},\ and\ \bibinfo {author} {\bibfnamefont {G.}~\bibnamefont {M\"oller}},\ }\bibfield  {title} {\bibinfo {title} {Particle entanglement spectra for quantum hall states on lattices},\ }\href {https://doi.org/10.1103/PhysRevB.86.165314} {\bibfield  {journal} {\bibinfo  {journal} {Phys. Rev. B}\ }\textbf {\bibinfo {volume} {86}},\ \bibinfo {pages} {165314} (\bibinfo {year} {2012})}\BibitemShut {NoStop}%
\bibitem [{\citenamefont {He}\ \emph {et~al.}(2015)\citenamefont {He}, \citenamefont {Bhattacharjee}, \citenamefont {Moessner},\ and\ \citenamefont {Pollmann}}]{He2015BIQH}%
  \BibitemOpen
  \bibfield  {author} {\bibinfo {author} {\bibfnamefont {Y.-C.}\ \bibnamefont {He}}, \bibinfo {author} {\bibfnamefont {S.}~\bibnamefont {Bhattacharjee}}, \bibinfo {author} {\bibfnamefont {R.}~\bibnamefont {Moessner}},\ and\ \bibinfo {author} {\bibfnamefont {F.}~\bibnamefont {Pollmann}},\ }\bibfield  {title} {\bibinfo {title} {Bosonic integer quantum hall effect in an interacting lattice model},\ }\href {https://doi.org/10.1103/PhysRevLett.115.116803} {\bibfield  {journal} {\bibinfo  {journal} {Phys. Rev. Lett.}\ }\textbf {\bibinfo {volume} {115}},\ \bibinfo {pages} {116803} (\bibinfo {year} {2015})}\BibitemShut {NoStop}%
\bibitem [{\citenamefont {Sterdyniak}\ \emph {et~al.}(2015)\citenamefont {Sterdyniak}, \citenamefont {Cooper},\ and\ \citenamefont {Regnault}}]{Sterdyniak2015BIQH}%
  \BibitemOpen
  \bibfield  {author} {\bibinfo {author} {\bibfnamefont {A.}~\bibnamefont {Sterdyniak}}, \bibinfo {author} {\bibfnamefont {N.~R.}\ \bibnamefont {Cooper}},\ and\ \bibinfo {author} {\bibfnamefont {N.}~\bibnamefont {Regnault}},\ }\bibfield  {title} {\bibinfo {title} {Bosonic integer quantum hall effect in optical flux lattices},\ }\href {https://doi.org/10.1103/PhysRevLett.115.116802} {\bibfield  {journal} {\bibinfo  {journal} {Phys. Rev. Lett.}\ }\textbf {\bibinfo {volume} {115}},\ \bibinfo {pages} {116802} (\bibinfo {year} {2015})}\BibitemShut {NoStop}%
\bibitem [{\citenamefont {He}\ \emph {et~al.}(2017)\citenamefont {He}, \citenamefont {Grusdt}, \citenamefont {Kaufman}, \citenamefont {Greiner},\ and\ \citenamefont {Vishwanath}}]{Vishvanath2017}%
  \BibitemOpen
  \bibfield  {author} {\bibinfo {author} {\bibfnamefont {Y.-C.}\ \bibnamefont {He}}, \bibinfo {author} {\bibfnamefont {F.}~\bibnamefont {Grusdt}}, \bibinfo {author} {\bibfnamefont {A.}~\bibnamefont {Kaufman}}, \bibinfo {author} {\bibfnamefont {M.}~\bibnamefont {Greiner}},\ and\ \bibinfo {author} {\bibfnamefont {A.}~\bibnamefont {Vishwanath}},\ }\bibfield  {title} {\bibinfo {title} {Realizing and adiabatically preparing bosonic integer and fractional quantum hall states in optical lattices},\ }\href {https://doi.org/10.1103/PhysRevB.96.201103} {\bibfield  {journal} {\bibinfo  {journal} {Phys. Rev. B}\ }\textbf {\bibinfo {volume} {96}},\ \bibinfo {pages} {201103} (\bibinfo {year} {2017})}\BibitemShut {NoStop}%
\bibitem [{\citenamefont {Rosson}\ \emph {et~al.}(2019)\citenamefont {Rosson}, \citenamefont {Lubasch}, \citenamefont {Kiffner},\ and\ \citenamefont {Jaksch}}]{Jaksch2019DMRG}%
  \BibitemOpen
  \bibfield  {author} {\bibinfo {author} {\bibfnamefont {P.}~\bibnamefont {Rosson}}, \bibinfo {author} {\bibfnamefont {M.}~\bibnamefont {Lubasch}}, \bibinfo {author} {\bibfnamefont {M.}~\bibnamefont {Kiffner}},\ and\ \bibinfo {author} {\bibfnamefont {D.}~\bibnamefont {Jaksch}},\ }\bibfield  {title} {\bibinfo {title} {Bosonic fractional quantum hall states on a finite cylinder},\ }\href {https://doi.org/10.1103/PhysRevA.99.033603} {\bibfield  {journal} {\bibinfo  {journal} {Phys. Rev. A}\ }\textbf {\bibinfo {volume} {99}},\ \bibinfo {pages} {033603} (\bibinfo {year} {2019})}\BibitemShut {NoStop}%
\bibitem [{\citenamefont {Palm}\ \emph {et~al.}(2021)\citenamefont {Palm}, \citenamefont {Buser}, \citenamefont {L\'eonard}, \citenamefont {Aidelsburger}, \citenamefont {Schollw\"ock},\ and\ \citenamefont {Grusdt}}]{Grusdt2021DMRG}%
  \BibitemOpen
  \bibfield  {author} {\bibinfo {author} {\bibfnamefont {F.~A.}\ \bibnamefont {Palm}}, \bibinfo {author} {\bibfnamefont {M.}~\bibnamefont {Buser}}, \bibinfo {author} {\bibfnamefont {J.}~\bibnamefont {L\'eonard}}, \bibinfo {author} {\bibfnamefont {M.}~\bibnamefont {Aidelsburger}}, \bibinfo {author} {\bibfnamefont {U.}~\bibnamefont {Schollw\"ock}},\ and\ \bibinfo {author} {\bibfnamefont {F.}~\bibnamefont {Grusdt}},\ }\bibfield  {title} {\bibinfo {title} {Bosonic pfaffian state in the hofstadter-bose-hubbard model},\ }\href {https://doi.org/10.1103/PhysRevB.103.L161101} {\bibfield  {journal} {\bibinfo  {journal} {Phys. Rev. B}\ }\textbf {\bibinfo {volume} {103}},\ \bibinfo {pages} {L161101} (\bibinfo {year} {2021})}\BibitemShut {NoStop}%
\bibitem [{\citenamefont {Boesl}\ \emph {et~al.}(2022)\citenamefont {Boesl}, \citenamefont {Dilip}, \citenamefont {Pollmann},\ and\ \citenamefont {Knap}}]{Boesl2022}%
  \BibitemOpen
  \bibfield  {author} {\bibinfo {author} {\bibfnamefont {J.}~\bibnamefont {Boesl}}, \bibinfo {author} {\bibfnamefont {R.}~\bibnamefont {Dilip}}, \bibinfo {author} {\bibfnamefont {F.}~\bibnamefont {Pollmann}},\ and\ \bibinfo {author} {\bibfnamefont {M.}~\bibnamefont {Knap}},\ }\bibfield  {title} {\bibinfo {title} {Characterizing fractional topological phases of lattice bosons near the first mott lobe},\ }\href {https://doi.org/10.1103/PhysRevB.105.075135} {\bibfield  {journal} {\bibinfo  {journal} {Phys. Rev. B}\ }\textbf {\bibinfo {volume} {105}},\ \bibinfo {pages} {075135} (\bibinfo {year} {2022})}\BibitemShut {NoStop}%
\bibitem [{\citenamefont {Gerster}\ \emph {et~al.}(2017)\citenamefont {Gerster}, \citenamefont {Rizzi}, \citenamefont {Silvi}, \citenamefont {Dalmonte},\ and\ \citenamefont {Montangero}}]{Rizzi2017TTN-FQH}%
  \BibitemOpen
  \bibfield  {author} {\bibinfo {author} {\bibfnamefont {M.}~\bibnamefont {Gerster}}, \bibinfo {author} {\bibfnamefont {M.}~\bibnamefont {Rizzi}}, \bibinfo {author} {\bibfnamefont {P.}~\bibnamefont {Silvi}}, \bibinfo {author} {\bibfnamefont {M.}~\bibnamefont {Dalmonte}},\ and\ \bibinfo {author} {\bibfnamefont {S.}~\bibnamefont {Montangero}},\ }\bibfield  {title} {\bibinfo {title} {Fractional quantum hall effect in the interacting hofstadter model via tensor networks},\ }\href {https://doi.org/10.1103/PhysRevB.96.195123} {\bibfield  {journal} {\bibinfo  {journal} {Phys. Rev. B}\ }\textbf {\bibinfo {volume} {96}},\ \bibinfo {pages} {195123} (\bibinfo {year} {2017})}\BibitemShut {NoStop}%
\bibitem [{\citenamefont {Macaluso}\ \emph {et~al.}(2020)\citenamefont {Macaluso}, \citenamefont {Comparin}, \citenamefont {Umucal\ifmmode \imath \else~\i \fi{}lar}, \citenamefont {Gerster}, \citenamefont {Montangero}, \citenamefont {Rizzi},\ and\ \citenamefont {Carusotto}}]{Rizzi2020}%
  \BibitemOpen
  \bibfield  {author} {\bibinfo {author} {\bibfnamefont {E.}~\bibnamefont {Macaluso}}, \bibinfo {author} {\bibfnamefont {T.}~\bibnamefont {Comparin}}, \bibinfo {author} {\bibfnamefont {R.~O.}\ \bibnamefont {Umucal\ifmmode \imath \else~\i \fi{}lar}}, \bibinfo {author} {\bibfnamefont {M.}~\bibnamefont {Gerster}}, \bibinfo {author} {\bibfnamefont {S.}~\bibnamefont {Montangero}}, \bibinfo {author} {\bibfnamefont {M.}~\bibnamefont {Rizzi}},\ and\ \bibinfo {author} {\bibfnamefont {I.}~\bibnamefont {Carusotto}},\ }\bibfield  {title} {\bibinfo {title} {Charge and statistics of lattice quasiholes from density measurements: A tree tensor network study},\ }\href {https://doi.org/10.1103/PhysRevResearch.2.013145} {\bibfield  {journal} {\bibinfo  {journal} {Phys. Rev. Res.}\ }\textbf {\bibinfo {volume} {2}},\ \bibinfo {pages} {013145} (\bibinfo {year} {2020})}\BibitemShut {NoStop}%
\bibitem [{\citenamefont {Jordan}\ \emph {et~al.}(2008)\citenamefont {Jordan}, \citenamefont {Or\'us}, \citenamefont {Vidal}, \citenamefont {Verstraete},\ and\ \citenamefont {Cirac}}]{Jordan2008iPEPS}%
  \BibitemOpen
  \bibfield  {author} {\bibinfo {author} {\bibfnamefont {J.}~\bibnamefont {Jordan}}, \bibinfo {author} {\bibfnamefont {R.}~\bibnamefont {Or\'us}}, \bibinfo {author} {\bibfnamefont {G.}~\bibnamefont {Vidal}}, \bibinfo {author} {\bibfnamefont {F.}~\bibnamefont {Verstraete}},\ and\ \bibinfo {author} {\bibfnamefont {J.~I.}\ \bibnamefont {Cirac}},\ }\bibfield  {title} {\bibinfo {title} {Classical simulation of infinite-size quantum lattice systems in two spatial dimensions},\ }\href {https://doi.org/10.1103/PhysRevLett.101.250602} {\bibfield  {journal} {\bibinfo  {journal} {Phys. Rev. Lett.}\ }\textbf {\bibinfo {volume} {101}},\ \bibinfo {pages} {250602} (\bibinfo {year} {2008})}\BibitemShut {NoStop}%
\bibitem [{\citenamefont {{Baxter}}(1968)}]{Baxter1968CTMRG1}%
  \BibitemOpen
  \bibfield  {author} {\bibinfo {author} {\bibfnamefont {R.~J.}\ \bibnamefont {{Baxter}}},\ }\bibfield  {title} {\bibinfo {title} {{Dimers on a Rectangular Lattice}},\ }\href {https://doi.org/10.1063/1.1664623} {\bibfield  {journal} {\bibinfo  {journal} {Journal of Mathematical Physics}\ }\textbf {\bibinfo {volume} {9}},\ \bibinfo {pages} {650} (\bibinfo {year} {1968})}\BibitemShut {NoStop}%
\bibitem [{\citenamefont {{Baxter}}(1978)}]{Baxter78CTMRG2}%
  \BibitemOpen
  \bibfield  {author} {\bibinfo {author} {\bibfnamefont {R.~J.}\ \bibnamefont {{Baxter}}},\ }\bibfield  {title} {\bibinfo {title} {{Variational approximations for square lattice models in statistical mechanics}},\ }\href {https://doi.org/10.1007/BF01011693} {\bibfield  {journal} {\bibinfo  {journal} {Journal of Statistical Physics}\ }\textbf {\bibinfo {volume} {19}},\ \bibinfo {pages} {461} (\bibinfo {year} {1978})}\BibitemShut {NoStop}%
\bibitem [{\citenamefont {Nishino}\ and\ \citenamefont {Okunishi}(1996)}]{Nishino96CTMRG}%
  \BibitemOpen
  \bibfield  {author} {\bibinfo {author} {\bibfnamefont {T.}~\bibnamefont {Nishino}}\ and\ \bibinfo {author} {\bibfnamefont {K.}~\bibnamefont {Okunishi}},\ }\bibfield  {title} {\bibinfo {title} {Corner transfer matrix renormalization group method},\ }\href {https://doi.org/10.1143/jpsj.65.891} {\bibfield  {journal} {\bibinfo  {journal} {Journal of the Physical Society of Japan}\ }\textbf {\bibinfo {volume} {65}},\ \bibinfo {pages} {891} (\bibinfo {year} {1996})}\BibitemShut {NoStop}%
\bibitem [{\citenamefont {Nishino}\ and\ \citenamefont {Okunishi}(1997)}]{Nishino97CTMRG}%
  \BibitemOpen
  \bibfield  {author} {\bibinfo {author} {\bibfnamefont {T.}~\bibnamefont {Nishino}}\ and\ \bibinfo {author} {\bibfnamefont {K.}~\bibnamefont {Okunishi}},\ }\bibfield  {title} {\bibinfo {title} {Corner transfer matrix algorithm for classical renormalization group},\ }\href {https://doi.org/10.1143/JPSJ.66.3040} {\bibfield  {journal} {\bibinfo  {journal} {Journal of the Physical Society of Japan}\ }\textbf {\bibinfo {volume} {66}},\ \bibinfo {pages} {3040} (\bibinfo {year} {1997})},\ \Eprint {https://arxiv.org/abs/https://doi.org/10.1143/JPSJ.66.3040} {https://doi.org/10.1143/JPSJ.66.3040} \BibitemShut {NoStop}%
\bibitem [{\citenamefont {Or\'us}\ and\ \citenamefont {Vidal}(2009)}]{Orus2009CTMRG}%
  \BibitemOpen
  \bibfield  {author} {\bibinfo {author} {\bibfnamefont {R.}~\bibnamefont {Or\'us}}\ and\ \bibinfo {author} {\bibfnamefont {G.}~\bibnamefont {Vidal}},\ }\bibfield  {title} {\bibinfo {title} {Simulation of two-dimensional quantum systems on an infinite lattice revisited: Corner transfer matrix for tensor contraction},\ }\href {https://doi.org/10.1103/PhysRevB.80.094403} {\bibfield  {journal} {\bibinfo  {journal} {Phys. Rev. B}\ }\textbf {\bibinfo {volume} {80}},\ \bibinfo {pages} {094403} (\bibinfo {year} {2009})}\BibitemShut {NoStop}%
\bibitem [{\citenamefont {Corboz}\ \emph {et~al.}(2010)\citenamefont {Corboz}, \citenamefont {Jordan},\ and\ \citenamefont {Vidal}}]{Corboz2010}%
  \BibitemOpen
  \bibfield  {author} {\bibinfo {author} {\bibfnamefont {P.}~\bibnamefont {Corboz}}, \bibinfo {author} {\bibfnamefont {J.}~\bibnamefont {Jordan}},\ and\ \bibinfo {author} {\bibfnamefont {G.}~\bibnamefont {Vidal}},\ }\bibfield  {title} {\bibinfo {title} {Simulation of fermionic lattice models in two dimensions with projected entangled-pair states: Next-nearest neighbor hamiltonians},\ }\href {https://doi.org/10.1103/PhysRevB.82.245119} {\bibfield  {journal} {\bibinfo  {journal} {Phys. Rev. B}\ }\textbf {\bibinfo {volume} {82}},\ \bibinfo {pages} {245119} (\bibinfo {year} {2010})}\BibitemShut {NoStop}%
\bibitem [{\citenamefont {Corboz}\ \emph {et~al.}(2011)\citenamefont {Corboz}, \citenamefont {White}, \citenamefont {Vidal},\ and\ \citenamefont {Troyer}}]{Corboz2011}%
  \BibitemOpen
  \bibfield  {author} {\bibinfo {author} {\bibfnamefont {P.}~\bibnamefont {Corboz}}, \bibinfo {author} {\bibfnamefont {S.~R.}\ \bibnamefont {White}}, \bibinfo {author} {\bibfnamefont {G.}~\bibnamefont {Vidal}},\ and\ \bibinfo {author} {\bibfnamefont {M.}~\bibnamefont {Troyer}},\ }\bibfield  {title} {\bibinfo {title} {Stripes in the two-dimensional $t$-$j$ model with infinite projected entangled-pair states},\ }\href {https://doi.org/10.1103/PhysRevB.84.041108} {\bibfield  {journal} {\bibinfo  {journal} {Phys. Rev. B}\ }\textbf {\bibinfo {volume} {84}},\ \bibinfo {pages} {041108} (\bibinfo {year} {2011})}\BibitemShut {NoStop}%
\bibitem [{\citenamefont {Corboz}\ \emph {et~al.}(2014)\citenamefont {Corboz}, \citenamefont {Rice},\ and\ \citenamefont {Troyer}}]{Corboz2014}%
  \BibitemOpen
  \bibfield  {author} {\bibinfo {author} {\bibfnamefont {P.}~\bibnamefont {Corboz}}, \bibinfo {author} {\bibfnamefont {T.~M.}\ \bibnamefont {Rice}},\ and\ \bibinfo {author} {\bibfnamefont {M.}~\bibnamefont {Troyer}},\ }\bibfield  {title} {\bibinfo {title} {Competing states in the $t$-$j$ model: Uniform $d$-wave state versus stripe state},\ }\href {https://doi.org/10.1103/PhysRevLett.113.046402} {\bibfield  {journal} {\bibinfo  {journal} {Phys. Rev. Lett.}\ }\textbf {\bibinfo {volume} {113}},\ \bibinfo {pages} {046402} (\bibinfo {year} {2014})}\BibitemShut {NoStop}%
\bibitem [{\citenamefont {Li}\ and\ \citenamefont {Haldane}(2008)}]{Li&Haldane2008}%
  \BibitemOpen
  \bibfield  {author} {\bibinfo {author} {\bibfnamefont {H.}~\bibnamefont {Li}}\ and\ \bibinfo {author} {\bibfnamefont {F.~D.~M.}\ \bibnamefont {Haldane}},\ }\bibfield  {title} {\bibinfo {title} {Entanglement spectrum as a generalization of entanglement entropy: Identification of topological order in non-abelian fractional quantum hall effect states},\ }\href {https://doi.org/10.1103/PhysRevLett.101.010504} {\bibfield  {journal} {\bibinfo  {journal} {Phys. Rev. Lett.}\ }\textbf {\bibinfo {volume} {101}},\ \bibinfo {pages} {010504} (\bibinfo {year} {2008})}\BibitemShut {NoStop}%
\bibitem [{\citenamefont {Haegeman}\ and\ \citenamefont {Verstraete}(2017)}]{Haegeman2017Medley}%
  \BibitemOpen
  \bibfield  {author} {\bibinfo {author} {\bibfnamefont {J.}~\bibnamefont {Haegeman}}\ and\ \bibinfo {author} {\bibfnamefont {F.}~\bibnamefont {Verstraete}},\ }\bibfield  {title} {\bibinfo {title} {Diagonalizing transfer matrices and matrix product operators: A medley of exact and computational methods},\ }\href {https://doi.org/10.1146/annurev-conmatphys-031016-025507} {\bibfield  {journal} {\bibinfo  {journal} {Annual Review of Condensed Matter Physics}\ }\textbf {\bibinfo {volume} {8}},\ \bibinfo {pages} {355} (\bibinfo {year} {2017})},\ \Eprint {https://arxiv.org/abs/https://doi.org/10.1146/annurev-conmatphys-031016-025507} {https://doi.org/10.1146/annurev-conmatphys-031016-025507} \BibitemShut {NoStop}%
\bibitem [{Sup()}]{SupplementaryMaterial}%
  \BibitemOpen
  \href@noop {} {}\bibinfo {note} {See Supplemental Material at (insert link here - currently below) for more details. The supplementary material additionally contains references \cite{hladnik1988spectrum, Zauner-Stauber2018VUMPS, Haegeman2013Excitations1, Heageman2019Tangentspacemethods, vandamme21ringsspectrum, ManyBodyQFTBookXGW}}\BibitemShut {NoStop}%
\bibitem [{\citenamefont {Corboz}(2016)}]{Corboz2016Variational}%
  \BibitemOpen
  \bibfield  {author} {\bibinfo {author} {\bibfnamefont {P.}~\bibnamefont {Corboz}},\ }\bibfield  {title} {\bibinfo {title} {Variational optimization with infinite projected entangled-pair states},\ }\href {https://doi.org/10.1103/PhysRevB.94.035133} {\bibfield  {journal} {\bibinfo  {journal} {Phys. Rev. B}\ }\textbf {\bibinfo {volume} {94}},\ \bibinfo {pages} {035133} (\bibinfo {year} {2016})}\BibitemShut {NoStop}%
\bibitem [{\citenamefont {Vanderstraeten}\ \emph {et~al.}(2016)\citenamefont {Vanderstraeten}, \citenamefont {Haegeman}, \citenamefont {Corboz},\ and\ \citenamefont {Verstraete}}]{Vanderstraeten2016Variational}%
  \BibitemOpen
  \bibfield  {author} {\bibinfo {author} {\bibfnamefont {L.}~\bibnamefont {Vanderstraeten}}, \bibinfo {author} {\bibfnamefont {J.}~\bibnamefont {Haegeman}}, \bibinfo {author} {\bibfnamefont {P.}~\bibnamefont {Corboz}},\ and\ \bibinfo {author} {\bibfnamefont {F.}~\bibnamefont {Verstraete}},\ }\bibfield  {title} {\bibinfo {title} {Gradient methods for variational optimization of projected entangled-pair states},\ }\href {https://doi.org/10.1103/PhysRevB.94.155123} {\bibfield  {journal} {\bibinfo  {journal} {Phys. Rev. B}\ }\textbf {\bibinfo {volume} {94}},\ \bibinfo {pages} {155123} (\bibinfo {year} {2016})}\BibitemShut {NoStop}%
\bibitem [{\citenamefont {Liao}\ \emph {et~al.}(2019)\citenamefont {Liao}, \citenamefont {Liu}, \citenamefont {Wang},\ and\ \citenamefont {Xiang}}]{Liao2019AD}%
  \BibitemOpen
  \bibfield  {author} {\bibinfo {author} {\bibfnamefont {H.-J.}\ \bibnamefont {Liao}}, \bibinfo {author} {\bibfnamefont {J.-G.}\ \bibnamefont {Liu}}, \bibinfo {author} {\bibfnamefont {L.}~\bibnamefont {Wang}},\ and\ \bibinfo {author} {\bibfnamefont {T.}~\bibnamefont {Xiang}},\ }\bibfield  {title} {\bibinfo {title} {Differentiable programming tensor networks},\ }\href {https://doi.org/10.1103/PhysRevX.9.031041} {\bibfield  {journal} {\bibinfo  {journal} {Phys. Rev. X}\ }\textbf {\bibinfo {volume} {9}},\ \bibinfo {pages} {031041} (\bibinfo {year} {2019})}\BibitemShut {NoStop}%
\bibitem [{\citenamefont {Naumann}\ \emph {et~al.}(2023)\citenamefont {Naumann}, \citenamefont {Weerda}, \citenamefont {Rizzi}, \citenamefont {Eisert},\ and\ \citenamefont {Schmoll}}]{varipeps}%
  \BibitemOpen
  \bibfield  {author} {\bibinfo {author} {\bibfnamefont {J.}~\bibnamefont {Naumann}}, \bibinfo {author} {\bibfnamefont {E.~L.}\ \bibnamefont {Weerda}}, \bibinfo {author} {\bibfnamefont {M.}~\bibnamefont {Rizzi}}, \bibinfo {author} {\bibfnamefont {J.}~\bibnamefont {Eisert}},\ and\ \bibinfo {author} {\bibfnamefont {P.}~\bibnamefont {Schmoll}},\ }\href@noop {} {\bibinfo {title} {varipeps -- a versatile tensor network library for variational ground state simulations in two spatial dimensions}} (\bibinfo {year} {2023}),\ \Eprint {https://arxiv.org/abs/2308.12358} {arXiv:2308.12358 [cond-mat.str-el]} \BibitemShut {NoStop}%
\bibitem [{\citenamefont {Hasik}\ \emph {et~al.}(2021)\citenamefont {Hasik}, \citenamefont {Poilblanc},\ and\ \citenamefont {Becca}}]{Hasik2021}%
  \BibitemOpen
  \bibfield  {author} {\bibinfo {author} {\bibfnamefont {J.}~\bibnamefont {Hasik}}, \bibinfo {author} {\bibfnamefont {D.}~\bibnamefont {Poilblanc}},\ and\ \bibinfo {author} {\bibfnamefont {F.}~\bibnamefont {Becca}},\ }\bibfield  {title} {\bibinfo {title} {{Investigation of the Néel phase of the frustrated Heisenberg antiferromagnet by differentiable symmetric tensor networks}},\ }\href {https://doi.org/10.21468/SciPostPhys.10.1.012} {\bibfield  {journal} {\bibinfo  {journal} {SciPost Phys.}\ }\textbf {\bibinfo {volume} {10}},\ \bibinfo {pages} {012} (\bibinfo {year} {2021})}\BibitemShut {NoStop}%
\bibitem [{\citenamefont {Hasik}\ \emph {et~al.}(2022{\natexlab{b}})\citenamefont {Hasik}, \citenamefont {Mbeng}, \citenamefont {Capponi}, \citenamefont {Becca},\ and\ \citenamefont {L\"auchli}}]{Hasik2022}%
  \BibitemOpen
  \bibfield  {author} {\bibinfo {author} {\bibfnamefont {J.}~\bibnamefont {Hasik}}, \bibinfo {author} {\bibfnamefont {G.~B.}\ \bibnamefont {Mbeng}}, \bibinfo {author} {\bibfnamefont {S.}~\bibnamefont {Capponi}}, \bibinfo {author} {\bibfnamefont {F.}~\bibnamefont {Becca}},\ and\ \bibinfo {author} {\bibfnamefont {A.~M.}\ \bibnamefont {L\"auchli}},\ }\bibfield  {title} {\bibinfo {title} {Symmetric projected entangled-pair states analysis of a phase transition in coupled spin-$\frac{1}{2}$ ladders},\ }\href {https://doi.org/10.1103/PhysRevB.106.125154} {\bibfield  {journal} {\bibinfo  {journal} {Phys. Rev. B}\ }\textbf {\bibinfo {volume} {106}},\ \bibinfo {pages} {125154} (\bibinfo {year} {2022}{\natexlab{b}})}\BibitemShut {NoStop}%
\bibitem [{\citenamefont {Tang}\ \emph {et~al.}(2011)\citenamefont {Tang}, \citenamefont {Mei},\ and\ \citenamefont {Wen}}]{Wen2011}%
  \BibitemOpen
  \bibfield  {author} {\bibinfo {author} {\bibfnamefont {E.}~\bibnamefont {Tang}}, \bibinfo {author} {\bibfnamefont {J.-W.}\ \bibnamefont {Mei}},\ and\ \bibinfo {author} {\bibfnamefont {X.-G.}\ \bibnamefont {Wen}},\ }\bibfield  {title} {\bibinfo {title} {High-temperature fractional quantum hall states},\ }\href {https://doi.org/10.1103/PhysRevLett.106.236802} {\bibfield  {journal} {\bibinfo  {journal} {Phys. Rev. Lett.}\ }\textbf {\bibinfo {volume} {106}},\ \bibinfo {pages} {236802} (\bibinfo {year} {2011})}\BibitemShut {NoStop}%
\bibitem [{\citenamefont {Neupert}\ \emph {et~al.}(2011)\citenamefont {Neupert}, \citenamefont {Santos}, \citenamefont {Chamon},\ and\ \citenamefont {Mudry}}]{Neupert2011}%
  \BibitemOpen
  \bibfield  {author} {\bibinfo {author} {\bibfnamefont {T.}~\bibnamefont {Neupert}}, \bibinfo {author} {\bibfnamefont {L.}~\bibnamefont {Santos}}, \bibinfo {author} {\bibfnamefont {C.}~\bibnamefont {Chamon}},\ and\ \bibinfo {author} {\bibfnamefont {C.}~\bibnamefont {Mudry}},\ }\bibfield  {title} {\bibinfo {title} {Fractional quantum hall states at zero magnetic field},\ }\href {https://doi.org/10.1103/PhysRevLett.106.236804} {\bibfield  {journal} {\bibinfo  {journal} {Phys. Rev. Lett.}\ }\textbf {\bibinfo {volume} {106}},\ \bibinfo {pages} {236804} (\bibinfo {year} {2011})}\BibitemShut {NoStop}%
\bibitem [{\citenamefont {Sun}\ \emph {et~al.}(2011)\citenamefont {Sun}, \citenamefont {Gu}, \citenamefont {Katsura},\ and\ \citenamefont {Das~Sarma}}]{Sun2011}%
  \BibitemOpen
  \bibfield  {author} {\bibinfo {author} {\bibfnamefont {K.}~\bibnamefont {Sun}}, \bibinfo {author} {\bibfnamefont {Z.}~\bibnamefont {Gu}}, \bibinfo {author} {\bibfnamefont {H.}~\bibnamefont {Katsura}},\ and\ \bibinfo {author} {\bibfnamefont {S.}~\bibnamefont {Das~Sarma}},\ }\bibfield  {title} {\bibinfo {title} {Nearly flatbands with nontrivial topology},\ }\href {https://doi.org/10.1103/PhysRevLett.106.236803} {\bibfield  {journal} {\bibinfo  {journal} {Phys. Rev. Lett.}\ }\textbf {\bibinfo {volume} {106}},\ \bibinfo {pages} {236803} (\bibinfo {year} {2011})}\BibitemShut {NoStop}%
\bibitem [{\citenamefont {Aidelsburger}(2018)}]{aidelsburger2018review}%
  \BibitemOpen
  \bibfield  {author} {\bibinfo {author} {\bibfnamefont {M.}~\bibnamefont {Aidelsburger}},\ }\bibfield  {title} {\bibinfo {title} {Artificial gauge fields and topology with ultracold atoms in optical lattices},\ }\href@noop {} {\bibfield  {journal} {\bibinfo  {journal} {Journal of Physics B: Atomic, Molecular and Optical Physics}\ }\textbf {\bibinfo {volume} {51}},\ \bibinfo {pages} {193001} (\bibinfo {year} {2018})}\BibitemShut {NoStop}%
\bibitem [{\citenamefont {Cirac}\ \emph {et~al.}(2011)\citenamefont {Cirac}, \citenamefont {Poilblanc}, \citenamefont {Schuch},\ and\ \citenamefont {Verstraete}}]{Cirac2011BBC}%
  \BibitemOpen
  \bibfield  {author} {\bibinfo {author} {\bibfnamefont {J.~I.}\ \bibnamefont {Cirac}}, \bibinfo {author} {\bibfnamefont {D.}~\bibnamefont {Poilblanc}}, \bibinfo {author} {\bibfnamefont {N.}~\bibnamefont {Schuch}},\ and\ \bibinfo {author} {\bibfnamefont {F.}~\bibnamefont {Verstraete}},\ }\bibfield  {title} {\bibinfo {title} {Entanglement spectrum and boundary theories with projected entangled-pair states},\ }\href {https://doi.org/10.1103/PhysRevB.83.245134} {\bibfield  {journal} {\bibinfo  {journal} {Phys. Rev. B}\ }\textbf {\bibinfo {volume} {83}},\ \bibinfo {pages} {245134} (\bibinfo {year} {2011})}\BibitemShut {NoStop}%
\bibitem [{\citenamefont {Jain}(1989)}]{Jain1989}%
  \BibitemOpen
  \bibfield  {author} {\bibinfo {author} {\bibfnamefont {J.~K.}\ \bibnamefont {Jain}},\ }\bibfield  {title} {\bibinfo {title} {Composite-fermion approach for the fractional quantum hall effect},\ }\href {https://doi.org/10.1103/PhysRevLett.63.199} {\bibfield  {journal} {\bibinfo  {journal} {Phys. Rev. Lett.}\ }\textbf {\bibinfo {volume} {63}},\ \bibinfo {pages} {199} (\bibinfo {year} {1989})}\BibitemShut {NoStop}%
\bibitem [{\citenamefont {Senthil}\ and\ \citenamefont {Levin}(2013)}]{Senthil2013BIQH}%
  \BibitemOpen
  \bibfield  {author} {\bibinfo {author} {\bibfnamefont {T.}~\bibnamefont {Senthil}}\ and\ \bibinfo {author} {\bibfnamefont {M.}~\bibnamefont {Levin}},\ }\bibfield  {title} {\bibinfo {title} {Integer quantum hall effect for bosons},\ }\href {https://doi.org/10.1103/PhysRevLett.110.046801} {\bibfield  {journal} {\bibinfo  {journal} {Phys. Rev. Lett.}\ }\textbf {\bibinfo {volume} {110}},\ \bibinfo {pages} {046801} (\bibinfo {year} {2013})}\BibitemShut {NoStop}%
\bibitem [{\citenamefont {Furukawa}\ and\ \citenamefont {Ueda}(2013)}]{Furukawa2013BIQH}%
  \BibitemOpen
  \bibfield  {author} {\bibinfo {author} {\bibfnamefont {S.}~\bibnamefont {Furukawa}}\ and\ \bibinfo {author} {\bibfnamefont {M.}~\bibnamefont {Ueda}},\ }\bibfield  {title} {\bibinfo {title} {Integer quantum hall state in two-component bose gases in a synthetic magnetic field},\ }\href {https://doi.org/10.1103/PhysRevLett.111.090401} {\bibfield  {journal} {\bibinfo  {journal} {Phys. Rev. Lett.}\ }\textbf {\bibinfo {volume} {111}},\ \bibinfo {pages} {090401} (\bibinfo {year} {2013})}\BibitemShut {NoStop}%
\bibitem [{\citenamefont {M\"oller}\ and\ \citenamefont {Cooper}(2015)}]{Moller2015}%
  \BibitemOpen
  \bibfield  {author} {\bibinfo {author} {\bibfnamefont {G.}~\bibnamefont {M\"oller}}\ and\ \bibinfo {author} {\bibfnamefont {N.~R.}\ \bibnamefont {Cooper}},\ }\bibfield  {title} {\bibinfo {title} {Fractional chern insulators in harper-hofstadter bands with higher chern number},\ }\href {https://doi.org/10.1103/PhysRevLett.115.126401} {\bibfield  {journal} {\bibinfo  {journal} {Phys. Rev. Lett.}\ }\textbf {\bibinfo {volume} {115}},\ \bibinfo {pages} {126401} (\bibinfo {year} {2015})}\BibitemShut {NoStop}%
\bibitem [{\citenamefont {Ponomarenko}\ \emph {et~al.}(2013)\citenamefont {Ponomarenko}, \citenamefont {Gorbachev}, \citenamefont {Yu}, \citenamefont {Elias}, \citenamefont {Jalil}, \citenamefont {Patel}, \citenamefont {Mishchenko}, \citenamefont {Mayorov}, \citenamefont {Woods}, \citenamefont {Wallbank} \emph {et~al.}}]{ponomarenko2013cloning}%
  \BibitemOpen
  \bibfield  {author} {\bibinfo {author} {\bibfnamefont {L.}~\bibnamefont {Ponomarenko}}, \bibinfo {author} {\bibfnamefont {R.}~\bibnamefont {Gorbachev}}, \bibinfo {author} {\bibfnamefont {G.}~\bibnamefont {Yu}}, \bibinfo {author} {\bibfnamefont {D.}~\bibnamefont {Elias}}, \bibinfo {author} {\bibfnamefont {R.}~\bibnamefont {Jalil}}, \bibinfo {author} {\bibfnamefont {A.}~\bibnamefont {Patel}}, \bibinfo {author} {\bibfnamefont {A.}~\bibnamefont {Mishchenko}}, \bibinfo {author} {\bibfnamefont {A.}~\bibnamefont {Mayorov}}, \bibinfo {author} {\bibfnamefont {C.}~\bibnamefont {Woods}}, \bibinfo {author} {\bibfnamefont {J.}~\bibnamefont {Wallbank}}, \emph {et~al.},\ }\bibfield  {title} {\bibinfo {title} {Cloning of dirac fermions in graphene superlattices},\ }\href@noop {} {\bibfield  {journal} {\bibinfo  {journal} {Nature}\ }\textbf {\bibinfo {volume} {497}},\ \bibinfo {pages} {594} (\bibinfo {year} {2013})}\BibitemShut {NoStop}%
\bibitem [{\citenamefont {Bergholtz}\ and\ \citenamefont {Liu}(2013)}]{bergholtz2013FCI}%
  \BibitemOpen
  \bibfield  {author} {\bibinfo {author} {\bibfnamefont {E.~J.}\ \bibnamefont {Bergholtz}}\ and\ \bibinfo {author} {\bibfnamefont {Z.}~\bibnamefont {Liu}},\ }\bibfield  {title} {\bibinfo {title} {Topological flat band models and fractional chern insulators},\ }\href@noop {} {\bibfield  {journal} {\bibinfo  {journal} {International Journal of Modern Physics B}\ }\textbf {\bibinfo {volume} {27}},\ \bibinfo {pages} {1330017} (\bibinfo {year} {2013})}\BibitemShut {NoStop}%
\bibitem [{\citenamefont {Zhang}\ \emph {et~al.}(2021)\citenamefont {Zhang}, \citenamefont {Sheng},\ and\ \citenamefont {Vishwanath}}]{Vishvanath2021}%
  \BibitemOpen
  \bibfield  {author} {\bibinfo {author} {\bibfnamefont {Y.-H.}\ \bibnamefont {Zhang}}, \bibinfo {author} {\bibfnamefont {D.~N.}\ \bibnamefont {Sheng}},\ and\ \bibinfo {author} {\bibfnamefont {A.}~\bibnamefont {Vishwanath}},\ }\bibfield  {title} {\bibinfo {title} {Su(4) chiral spin liquid, exciton supersolid, and electric detection in moir\'e bilayers},\ }\href {https://doi.org/10.1103/PhysRevLett.127.247701} {\bibfield  {journal} {\bibinfo  {journal} {Phys. Rev. Lett.}\ }\textbf {\bibinfo {volume} {127}},\ \bibinfo {pages} {247701} (\bibinfo {year} {2021})}\BibitemShut {NoStop}%
\bibitem [{\citenamefont {Wu}\ \emph {et~al.}(2023)\citenamefont {Wu}, \citenamefont {Wu},\ and\ \citenamefont {Yao}}]{Wu2023}%
  \BibitemOpen
  \bibfield  {author} {\bibinfo {author} {\bibfnamefont {Y.-M.}\ \bibnamefont {Wu}}, \bibinfo {author} {\bibfnamefont {Z.}~\bibnamefont {Wu}},\ and\ \bibinfo {author} {\bibfnamefont {H.}~\bibnamefont {Yao}},\ }\bibfield  {title} {\bibinfo {title} {Pair-density-wave and chiral superconductivity in twisted bilayer transition metal dichalcogenides},\ }\href {https://doi.org/10.1103/PhysRevLett.130.126001} {\bibfield  {journal} {\bibinfo  {journal} {Phys. Rev. Lett.}\ }\textbf {\bibinfo {volume} {130}},\ \bibinfo {pages} {126001} (\bibinfo {year} {2023})}\BibitemShut {NoStop}%
\bibitem [{\citenamefont {Parameswaran}\ \emph {et~al.}(2013{\natexlab{b}})\citenamefont {Parameswaran}, \citenamefont {Roy},\ and\ \citenamefont {Sondhi}}]{Parameswaran2013fractional}%
  \BibitemOpen
  \bibfield  {author} {\bibinfo {author} {\bibfnamefont {S.~A.}\ \bibnamefont {Parameswaran}}, \bibinfo {author} {\bibfnamefont {R.}~\bibnamefont {Roy}},\ and\ \bibinfo {author} {\bibfnamefont {S.~L.}\ \bibnamefont {Sondhi}},\ }\bibfield  {title} {\bibinfo {title} {Fractional quantum hall physics in topological flat bands},\ }\href@noop {} {\bibfield  {journal} {\bibinfo  {journal} {Comptes Rendus Physique}\ }\textbf {\bibinfo {volume} {14}},\ \bibinfo {pages} {816} (\bibinfo {year} {2013}{\natexlab{b}})}\BibitemShut {NoStop}%
\bibitem [{\citenamefont {Wu}\ \emph {et~al.}(2018)\citenamefont {Wu}, \citenamefont {Lovorn}, \citenamefont {Tutuc},\ and\ \citenamefont {MacDonald}}]{MacDonald2018}%
  \BibitemOpen
  \bibfield  {author} {\bibinfo {author} {\bibfnamefont {F.}~\bibnamefont {Wu}}, \bibinfo {author} {\bibfnamefont {T.}~\bibnamefont {Lovorn}}, \bibinfo {author} {\bibfnamefont {E.}~\bibnamefont {Tutuc}},\ and\ \bibinfo {author} {\bibfnamefont {A.~H.}\ \bibnamefont {MacDonald}},\ }\bibfield  {title} {\bibinfo {title} {Hubbard model physics in transition metal dichalcogenide moir\'e bands},\ }\href {https://doi.org/10.1103/PhysRevLett.121.026402} {\bibfield  {journal} {\bibinfo  {journal} {Phys. Rev. Lett.}\ }\textbf {\bibinfo {volume} {121}},\ \bibinfo {pages} {026402} (\bibinfo {year} {2018})}\BibitemShut {NoStop}%
\bibitem [{\citenamefont {Wang}\ \emph {et~al.}(2020)\citenamefont {Wang}, \citenamefont {Shih}, \citenamefont {Ghiotto}, \citenamefont {Xian}, \citenamefont {Rhodes}, \citenamefont {Tan}, \citenamefont {Claassen}, \citenamefont {Kennes}, \citenamefont {Bai}, \citenamefont {Kim}, \citenamefont {Watanabe}, \citenamefont {Taniguchi}, \citenamefont {Zhu}, \citenamefont {Hone}, \citenamefont {Rubio}, \citenamefont {Pasupathy},\ and\ \citenamefont {Dean}}]{Wang2020}%
  \BibitemOpen
  \bibfield  {author} {\bibinfo {author} {\bibfnamefont {L.}~\bibnamefont {Wang}}, \bibinfo {author} {\bibfnamefont {E.-M.}\ \bibnamefont {Shih}}, \bibinfo {author} {\bibfnamefont {A.}~\bibnamefont {Ghiotto}}, \bibinfo {author} {\bibfnamefont {L.}~\bibnamefont {Xian}}, \bibinfo {author} {\bibfnamefont {D.~A.}\ \bibnamefont {Rhodes}}, \bibinfo {author} {\bibfnamefont {C.}~\bibnamefont {Tan}}, \bibinfo {author} {\bibfnamefont {M.}~\bibnamefont {Claassen}}, \bibinfo {author} {\bibfnamefont {D.~M.}\ \bibnamefont {Kennes}}, \bibinfo {author} {\bibfnamefont {Y.}~\bibnamefont {Bai}}, \bibinfo {author} {\bibfnamefont {B.}~\bibnamefont {Kim}}, \bibinfo {author} {\bibfnamefont {K.}~\bibnamefont {Watanabe}}, \bibinfo {author} {\bibfnamefont {T.}~\bibnamefont {Taniguchi}}, \bibinfo {author} {\bibfnamefont {X.}~\bibnamefont {Zhu}}, \bibinfo {author} {\bibfnamefont {J.}~\bibnamefont {Hone}}, \bibinfo {author} {\bibfnamefont {A.}~\bibnamefont {Rubio}}, \bibinfo {author} {\bibfnamefont {A.~N.}\ \bibnamefont {Pasupathy}},\
  and\ \bibinfo {author} {\bibfnamefont {C.~R.}\ \bibnamefont {Dean}},\ }\bibfield  {title} {\bibinfo {title} {Correlated electronic phases in twisted bilayer transition metal dichalcogenides},\ }\href {https://doi.org/10.1038/s41563-020-0708-6} {\bibfield  {journal} {\bibinfo  {journal} {Nature Materials}\ }\textbf {\bibinfo {volume} {19}},\ \bibinfo {pages} {861} (\bibinfo {year} {2020})}\BibitemShut {NoStop}%
\bibitem [{\citenamefont {Kiese}\ \emph {et~al.}(2022)\citenamefont {Kiese}, \citenamefont {He}, \citenamefont {Hickey}, \citenamefont {Rubio},\ and\ \citenamefont {Kennes}}]{kiese2022tmds}%
  \BibitemOpen
  \bibfield  {author} {\bibinfo {author} {\bibfnamefont {D.}~\bibnamefont {Kiese}}, \bibinfo {author} {\bibfnamefont {Y.}~\bibnamefont {He}}, \bibinfo {author} {\bibfnamefont {C.}~\bibnamefont {Hickey}}, \bibinfo {author} {\bibfnamefont {A.}~\bibnamefont {Rubio}},\ and\ \bibinfo {author} {\bibfnamefont {D.~M.}\ \bibnamefont {Kennes}},\ }\bibfield  {title} {\bibinfo {title} {Tmds as a platform for spin liquid physics: A strong coupling study of twisted bilayer wse2},\ }\href@noop {} {\bibfield  {journal} {\bibinfo  {journal} {APL Materials}\ }\textbf {\bibinfo {volume} {10}} (\bibinfo {year} {2022})}\BibitemShut {NoStop}%
\bibitem [{\citenamefont {Ponsioen}\ \emph {et~al.}(2022)\citenamefont {Ponsioen}, \citenamefont {Assaad},\ and\ \citenamefont {Corboz}}]{Ponsioen22spectalfunctions}%
  \BibitemOpen
  \bibfield  {author} {\bibinfo {author} {\bibfnamefont {B.}~\bibnamefont {Ponsioen}}, \bibinfo {author} {\bibfnamefont {F.~F.}\ \bibnamefont {Assaad}},\ and\ \bibinfo {author} {\bibfnamefont {P.}~\bibnamefont {Corboz}},\ }\bibfield  {title} {\bibinfo {title} {{Automatic differentiation applied to excitations with projected entangled pair states}},\ }\href {https://doi.org/10.21468/SciPostPhys.12.1.006} {\bibfield  {journal} {\bibinfo  {journal} {SciPost Phys.}\ }\textbf {\bibinfo {volume} {12}},\ \bibinfo {pages} {006} (\bibinfo {year} {2022})}\BibitemShut {NoStop}%
\bibitem [{\citenamefont {Ponsioen}\ \emph {et~al.}(2023)\citenamefont {Ponsioen}, \citenamefont {Hasik},\ and\ \citenamefont {Corboz}}]{ponsioen2023improvedSF}%
  \BibitemOpen
  \bibfield  {author} {\bibinfo {author} {\bibfnamefont {B.}~\bibnamefont {Ponsioen}}, \bibinfo {author} {\bibfnamefont {J.}~\bibnamefont {Hasik}},\ and\ \bibinfo {author} {\bibfnamefont {P.}~\bibnamefont {Corboz}},\ }\href@noop {} {\bibinfo {title} {Improved summations of $n$-point correlation functions of projected entangled-pair states}} (\bibinfo {year} {2023}),\ \Eprint {https://arxiv.org/abs/2306.13327} {arXiv:2306.13327 [cond-mat.str-el]} \BibitemShut {NoStop}%
\bibitem [{\citenamefont {Tu}\ \emph {et~al.}(2023)\citenamefont {Tu}, \citenamefont {Vanderstraeten}, \citenamefont {Schuch}, \citenamefont {Lee}, \citenamefont {Kawashima},\ and\ \citenamefont {Chen}}]{tu2023generating}%
  \BibitemOpen
  \bibfield  {author} {\bibinfo {author} {\bibfnamefont {W.-L.}\ \bibnamefont {Tu}}, \bibinfo {author} {\bibfnamefont {L.}~\bibnamefont {Vanderstraeten}}, \bibinfo {author} {\bibfnamefont {N.}~\bibnamefont {Schuch}}, \bibinfo {author} {\bibfnamefont {H.-Y.}\ \bibnamefont {Lee}}, \bibinfo {author} {\bibfnamefont {N.}~\bibnamefont {Kawashima}},\ and\ \bibinfo {author} {\bibfnamefont {J.-Y.}\ \bibnamefont {Chen}},\ }\href@noop {} {\bibinfo {title} {Generating function for projected entangled-pair states}} (\bibinfo {year} {2023}),\ \Eprint {https://arxiv.org/abs/2307.08083} {arXiv:2307.08083 [cond-mat.str-el]} \BibitemShut {NoStop}%
\bibitem [{\citenamefont {Chi}\ \emph {et~al.}(2022)\citenamefont {Chi}, \citenamefont {Liu}, \citenamefont {Wan}, \citenamefont {Liao},\ and\ \citenamefont {Xiang}}]{Heisenberg22spectrum}%
  \BibitemOpen
  \bibfield  {author} {\bibinfo {author} {\bibfnamefont {R.}~\bibnamefont {Chi}}, \bibinfo {author} {\bibfnamefont {Y.}~\bibnamefont {Liu}}, \bibinfo {author} {\bibfnamefont {Y.}~\bibnamefont {Wan}}, \bibinfo {author} {\bibfnamefont {H.-J.}\ \bibnamefont {Liao}},\ and\ \bibinfo {author} {\bibfnamefont {T.}~\bibnamefont {Xiang}},\ }\bibfield  {title} {\bibinfo {title} {Spin excitation spectra of anisotropic spin-$1/2$ triangular lattice heisenberg antiferromagnets},\ }\href {https://doi.org/10.1103/PhysRevLett.129.227201} {\bibfield  {journal} {\bibinfo  {journal} {Phys. Rev. Lett.}\ }\textbf {\bibinfo {volume} {129}},\ \bibinfo {pages} {227201} (\bibinfo {year} {2022})}\BibitemShut {NoStop}%
\bibitem [{\citenamefont {Centre}(2021)}]{JURECA2021}%
  \BibitemOpen
  \bibfield  {author} {\bibinfo {author} {\bibfnamefont {J.~S.}\ \bibnamefont {Centre}},\ }\bibfield  {title} {\bibinfo {title} {Jureca: Data centric and booster modules implementing the modular supercomputing architecture at j{\"u}lich supercomputing centre},\ }\href {https://doi.org/10.17815/jlsrf-7-182} {\bibfield  {journal} {\bibinfo  {journal} {Journal of large-scale research facilities}\ }\textbf {\bibinfo {volume} {7}},\ \bibinfo {pages} {A182} (\bibinfo {year} {2021})}\BibitemShut {NoStop}%
\bibitem [{\citenamefont {Hladnik}\ and\ \citenamefont {Omladi{\v{c}}}(1988)}]{hladnik1988spectrum}%
  \BibitemOpen
  \bibfield  {author} {\bibinfo {author} {\bibfnamefont {M.}~\bibnamefont {Hladnik}}\ and\ \bibinfo {author} {\bibfnamefont {M.}~\bibnamefont {Omladi{\v{c}}}},\ }\bibfield  {title} {\bibinfo {title} {Spectrum of the product of operators},\ }\href@noop {} {\bibfield  {journal} {\bibinfo  {journal} {Proceedings of the American Mathematical Society}\ }\textbf {\bibinfo {volume} {102}},\ \bibinfo {pages} {300} (\bibinfo {year} {1988})}\BibitemShut {NoStop}%
\bibitem [{\citenamefont {Zauner-Stauber}\ \emph {et~al.}(2018)\citenamefont {Zauner-Stauber}, \citenamefont {Vanderstraeten}, \citenamefont {Fishman}, \citenamefont {Verstraete},\ and\ \citenamefont {Haegeman}}]{Zauner-Stauber2018VUMPS}%
  \BibitemOpen
  \bibfield  {author} {\bibinfo {author} {\bibfnamefont {V.}~\bibnamefont {Zauner-Stauber}}, \bibinfo {author} {\bibfnamefont {L.}~\bibnamefont {Vanderstraeten}}, \bibinfo {author} {\bibfnamefont {M.~T.}\ \bibnamefont {Fishman}}, \bibinfo {author} {\bibfnamefont {F.}~\bibnamefont {Verstraete}},\ and\ \bibinfo {author} {\bibfnamefont {J.}~\bibnamefont {Haegeman}},\ }\bibfield  {title} {\bibinfo {title} {Variational optimization algorithms for uniform matrix product states},\ }\href {https://doi.org/10.1103/PhysRevB.97.045145} {\bibfield  {journal} {\bibinfo  {journal} {Phys. Rev. B}\ }\textbf {\bibinfo {volume} {97}},\ \bibinfo {pages} {045145} (\bibinfo {year} {2018})}\BibitemShut {NoStop}%
\bibitem [{\citenamefont {Haegeman}\ \emph {et~al.}(2013)\citenamefont {Haegeman}, \citenamefont {Michalakis}, \citenamefont {Nachtergaele}, \citenamefont {Osborne}, \citenamefont {Schuch},\ and\ \citenamefont {Verstraete}}]{Haegeman2013Excitations1}%
  \BibitemOpen
  \bibfield  {author} {\bibinfo {author} {\bibfnamefont {J.}~\bibnamefont {Haegeman}}, \bibinfo {author} {\bibfnamefont {S.}~\bibnamefont {Michalakis}}, \bibinfo {author} {\bibfnamefont {B.}~\bibnamefont {Nachtergaele}}, \bibinfo {author} {\bibfnamefont {T.~J.}\ \bibnamefont {Osborne}}, \bibinfo {author} {\bibfnamefont {N.}~\bibnamefont {Schuch}},\ and\ \bibinfo {author} {\bibfnamefont {F.}~\bibnamefont {Verstraete}},\ }\bibfield  {title} {\bibinfo {title} {Elementary excitations in gapped quantum spin systems},\ }\href {https://doi.org/10.1103/PhysRevLett.111.080401} {\bibfield  {journal} {\bibinfo  {journal} {Phys. Rev. Lett.}\ }\textbf {\bibinfo {volume} {111}},\ \bibinfo {pages} {080401} (\bibinfo {year} {2013})}\BibitemShut {NoStop}%
\bibitem [{\citenamefont {Vanderstraeten}\ \emph {et~al.}(2019)\citenamefont {Vanderstraeten}, \citenamefont {Haegeman},\ and\ \citenamefont {Verstraete}}]{Heageman2019Tangentspacemethods}%
  \BibitemOpen
  \bibfield  {author} {\bibinfo {author} {\bibfnamefont {L.}~\bibnamefont {Vanderstraeten}}, \bibinfo {author} {\bibfnamefont {J.}~\bibnamefont {Haegeman}},\ and\ \bibinfo {author} {\bibfnamefont {F.}~\bibnamefont {Verstraete}},\ }\bibfield  {title} {\bibinfo {title} {{Tangent-space methods for uniform matrix product states}},\ }\href {https://doi.org/10.21468/SciPostPhysLectNotes.7} {\bibfield  {journal} {\bibinfo  {journal} {SciPost Phys. Lect. Notes}\ ,\ \bibinfo {pages} {7}} (\bibinfo {year} {2019})}\BibitemShut {NoStop}%
\bibitem [{\citenamefont {Van~Damme}\ \emph {et~al.}(2021)\citenamefont {Van~Damme}, \citenamefont {Vanhove}, \citenamefont {Haegeman}, \citenamefont {Verstraete},\ and\ \citenamefont {Vanderstraeten}}]{vandamme21ringsspectrum}%
  \BibitemOpen
  \bibfield  {author} {\bibinfo {author} {\bibfnamefont {M.}~\bibnamefont {Van~Damme}}, \bibinfo {author} {\bibfnamefont {R.}~\bibnamefont {Vanhove}}, \bibinfo {author} {\bibfnamefont {J.}~\bibnamefont {Haegeman}}, \bibinfo {author} {\bibfnamefont {F.}~\bibnamefont {Verstraete}},\ and\ \bibinfo {author} {\bibfnamefont {L.}~\bibnamefont {Vanderstraeten}},\ }\bibfield  {title} {\bibinfo {title} {Efficient matrix product state methods for extracting spectral information on rings and cylinders},\ }\href {https://doi.org/10.1103/PhysRevB.104.115142} {\bibfield  {journal} {\bibinfo  {journal} {Phys. Rev. B}\ }\textbf {\bibinfo {volume} {104}},\ \bibinfo {pages} {115142} (\bibinfo {year} {2021})}\BibitemShut {NoStop}%
\bibitem [{\citenamefont {Gang}(2007)}]{ManyBodyQFTBookXGW}%
  \BibitemOpen
  \bibfield  {author} {\bibinfo {author} {\bibfnamefont {W.~X.}\ \bibnamefont {Gang}},\ }\href {https://doi.org/10.1093/acprof:oso/9780199227259.001.0001} {\emph {\bibinfo {title} {{Quantum field theory of many-body systems: from the origin of sound to an origin of light and electrons}}}}\ (\bibinfo  {publisher} {Oxford University Press},\ \bibinfo {address} {Oxford},\ \bibinfo {year} {2007})\BibitemShut {NoStop}%
\bibitem [{\citenamefont {Nielsen}\ and\ \citenamefont {Ninomiya}(1981)}]{NIELSEN1981}%
  \BibitemOpen
  \bibfield  {author} {\bibinfo {author} {\bibfnamefont {H.}~\bibnamefont {Nielsen}}\ and\ \bibinfo {author} {\bibfnamefont {M.}~\bibnamefont {Ninomiya}},\ }\bibfield  {title} {\bibinfo {title} {Absence of neutrinos on a lattice: (i). proof by homotopy theory},\ }\href {https://doi.org/https://doi.org/10.1016/0550-3213(81)90361-8} {\bibfield  {journal} {\bibinfo  {journal} {Nuclear Physics B}\ }\textbf {\bibinfo {volume} {185}},\ \bibinfo {pages} {20} (\bibinfo {year} {1981})}\BibitemShut {NoStop}%
\end{thebibliography}%

\title{Supplementary Material
}

\maketitle

\section{Gauge and unit cell choices}
For treating the Harper-Hofstadter model using hard-core bosons at a flux per plaquette of $\phi = \pi/2$ we used the conventional Landau gauge choice. 
This leads to a $4 \times 1$ magnetic unit cell. 
For the ground state search we applied both a $4 \times 1$ and an enlarged $4 \times 4$ unit cell.
While both unit cell choices lead to the emergence of the Laughlin plateau, their use turns out to be convenient for different purposes.
The enlarged unit cell ensures a more stable and less biased optimization, albeit slower, and is therefore the choice for scanning the different filling regimes. The original magnetic unit cell offers instead the advantage of a full periodicity of the iPEPS in one direction, which is instrumental for the calculation of the entanglement spectra (see next section), and can be safely used within the stable plateau.
The same holds for the calculations of the soft-core bosons (with large interactions) at flux per plaquette $\phi = \pi/2$.

\begin{figure}[h]
    \centering
    \includegraphics[width=0.3\textwidth]{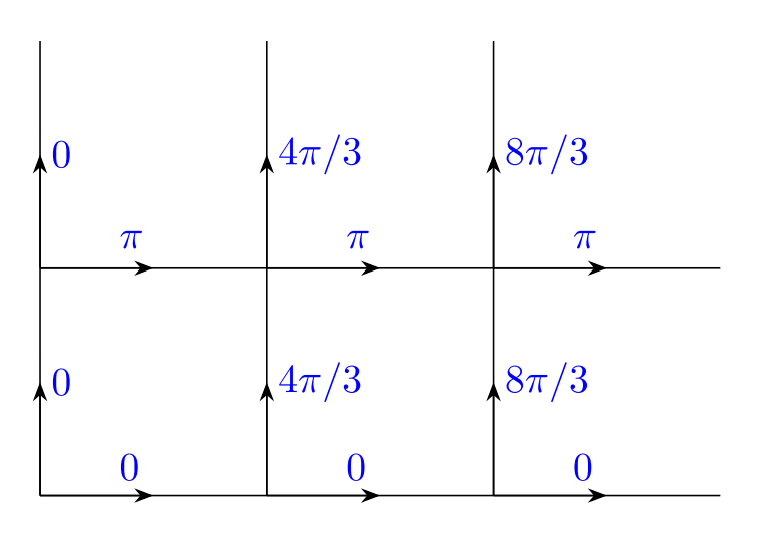}
    \caption{\raggedright \small Illustration of the custom gauge choice. The blue numbers illustrate a choice for the the value of gauge fields on the edges. If we consider a single plaquette, and add up all the values for the gauge fields they add to $\pi/3 \mod 2\pi$. Note that we must take a minus sign for those gauge fields corresponding to arrows that point clockwise for the plaquette in question.}
    \label{fig:gauge_choice}
\end{figure}

For the case of soft-core bosons at flux per plaquette of $\phi = \pi/3$ we choose a again can choose a Landau gauge and choose an enlarged unit cell of the iPEPS of $6 \times 6$. Here we observe the emergence of the incompressible plateaus shown in the main text.
We additionally use a custom gauge choice with a magnetic unit cell of $3 \times 2$ corresponding to the same value of the flux per plaquette: by using this as iPEPS unit cell, we are able to confirm the emergence of all plateaus with faster convergence.
The custom gauge choice is illustrated in in Fig. \ref{fig:gauge_choice}.
We note in addition that we did not to enforce local symmetries on the tensors in our calculation as it was done in previous works \cite{Hasik2022CSL}.

\section{Calculation of the low-lying Entanglement spectrum in the thermodynamic limit}
\begin{figure}[t]
    \centering
    \includegraphics[width=0.43\textwidth]{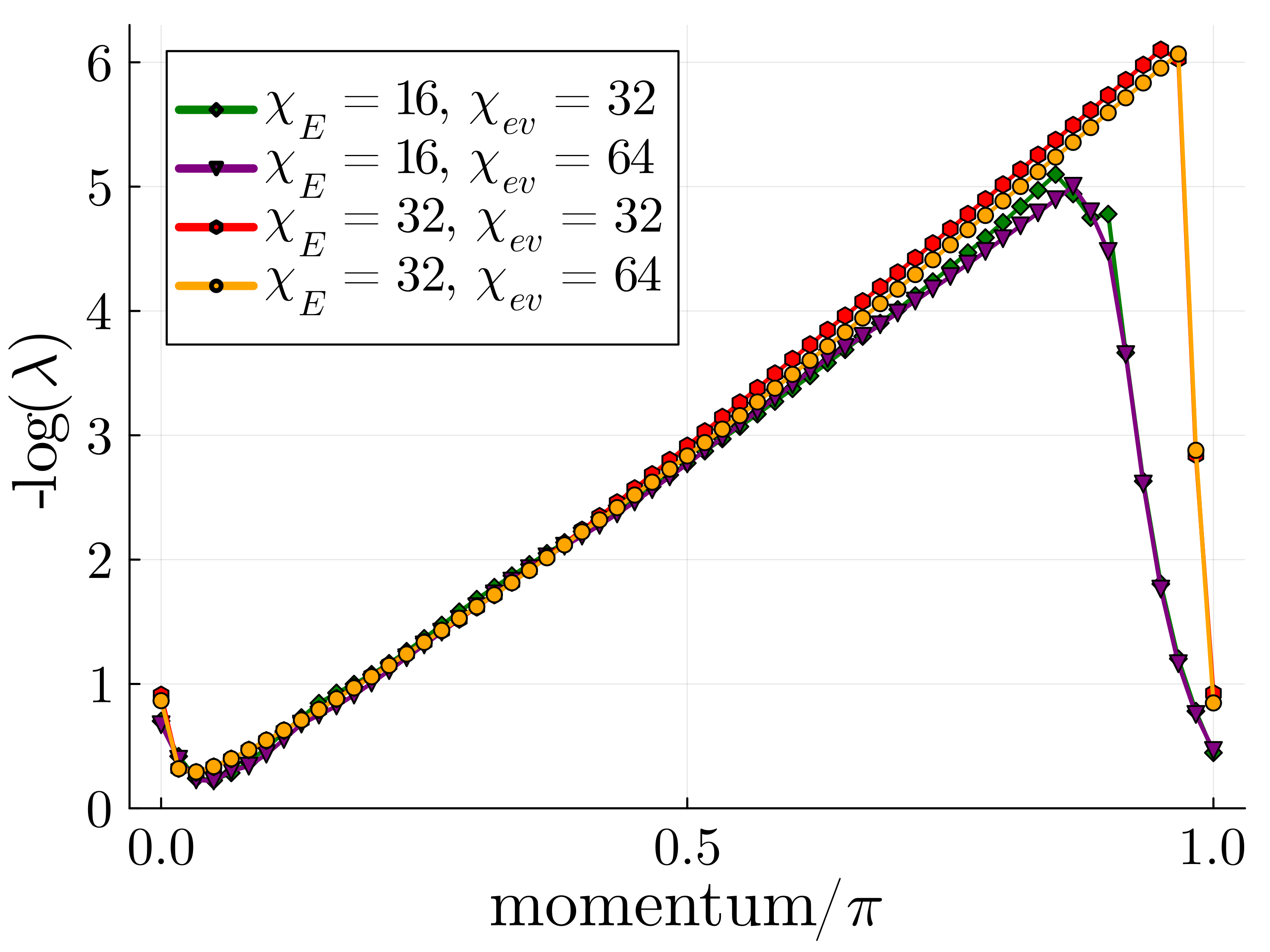}
    \caption{\raggedright \small Illustration showing that the bond dimension $\chi_{ev}$ used for the MPS approximation of the leading eigenvector in the entanglement spectrum calculation does not qualitatively change the spectrum, while value of the bond dimension used for the MPO approximation of the boundary Hamiltonian indeed increases the chirality of the spectrum.}
    \label{fig:chi_ev_comparison}
\end{figure}

In order to characterize the edge modes of a PEPS state we got from our variational optimization procedure, we can calculate the entanglement spectrum (ES). Following Li and Haldane \cite{Li&Haldane2008}, this is corresponding to the spectrum of the edge modes of the system. The entanglement spectrum is defined as the spectrum of the entanglement Hamiltonian $H_{ent}$, that in turn can identified from the reduced density matrix $\rho_l$ of a bipartition of the system $\rho_l = e^{-H_{ent}}$.
We can access the ES for our system by utilizing the bulk-boundary correspondence for PEPS \cite{Cirac2011BBC}, and can do so directly in the thermodynamic limit \cite{Haegeman2017Medley}.

Given a PEPS network, we bipartition it into a left and right side network.
These two networks have open physical indices, as well as open virtual indices at the boundary defined by the bipartition.
Tracing out all physical indices of this left (right) part of the PEPS network we obtain $\sigma_l$ ($\sigma_r$). We can think of $\sigma_l$ ($\sigma_r$) as the reduced density matrices of the virtual indices at the boundary for the left and right half of the original PEPS. The PEPS bulk-boundary correspondence allows us to express the reduced density matrix $\rho_l$ of the left-side physical indices 
in terms of these objects:
\begin{equation}
    \rho_l = U \sqrt{\sigma_l^T}\sigma_r \sqrt{\sigma_l^T}U^\dagger \, ,
\end{equation}
where $U$ is an isometry that preserves the spectrum. 
With the CTMRG procedure, we construct environment tensors that can be used to approximate $\sigma_l$ and $\sigma_r$. 
Specifically, the CTMRG we produces environment tensors with bond dimension $\chi_E$, that approximate semi-infinite stripes of a double layer PEPS network. Thus a product of these environment tensors can be used gives an bond dimension $\chi_E$ MPO-approximation of $\sigma_l$ or $\sigma_r$, as follows:
\begin{equation}
    \sigma_r \approx \begin{gathered}
\includegraphics[height=3cm]{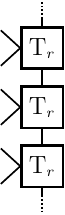}
\end{gathered} \hspace{0.2cm}, \hspace{0.6cm} \begin{gathered}
\includegraphics[height=0.85cm]{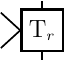}
\end{gathered} \approx \begin{gathered}
\includegraphics[height=1.05cm]{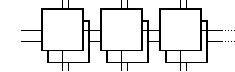}
\hspace{0.2cm}
\end{gathered}.
\end{equation}

For practical purposes we note that the operator $\sqrt{\sigma_l^T}\sigma_r \sqrt{\sigma_l^T}$ has the same eigenvalues as the operator $\sigma_l\sigma_r$: see Ref.~\cite{hladnik1988spectrum} for mathematical details. 
Note that therefore the approximation parameter from the main text is related to the MPO-bond dimension $\vartheta = \chi_E^2$. 
Therefore, to calculate the entanglement spectrum, we only need to find the eigenvalues of the product of the matrix product operators (MPOs) for $\sigma_l$ and $\sigma_r$, which is an MPO itself.
We obtain the lowest part of the ES by first using the VUMPS algorithm to approximate the leading eigenvector of $\sigma_l \sigma_r$ \cite{Zauner-Stauber2018VUMPS}. 
Starting from this leading eigenvector we use the quasi-particle excitation ansatz to find the energies of the low lying excitations at a given momentum \cite{Haegeman2013Excitations1, Heageman2019Tangentspacemethods}. 
The bond dimension $\chi_{ev}$ we choose for this leading eigenvector needs to be viewed as an additional approximation to be controlled. 
To this end, we note that a change in the bond dimension of this leading eigenvalue $\chi_{ev}$ does not qualitatively change the chiral character of the entanglement spectrum of the fractional Hall states treated in this work, but can lead to slightly more fluctuations in the spectral values, as illustrated in fig. \ref{fig:chi_ev_comparison}.
In the main text we show the result of these ES calculation and highlight especially the bond dimension of the MPO, that we use for the approximation of $\rho_l$ and hence $H_{ent}$, serves as a crucial approximation parameter, as it can be used to increase the chirality of the spectrum. This can be understood from the fact that the MPO bond dimension can be seen as controlling the length scale of interactions of the boundary Hamiltonian following the PEPS bulk-boundary correspondence. For a truly chiral entanglement spectrum, and hence boundary modes, it can be argued by analogy to \cite{NIELSEN1981}, that a truly non-local boundary Hamiltonian is needed \cite{Hasik2022CSL} which is only archived at infinite MPO bond dimension.
We further note in this context that for different Hamiltonians that permit additional symmetries such that $\sigma_l = \sigma_r$, one only needs to calculate the spectrum of one of these MPOs to get the spectrum of $\rho_l$.
\section{Entanglement spectrum on a cylinder of finite circumference}
\begin{figure}[t]
    \centering
    \includegraphics[width=0.43\textwidth]{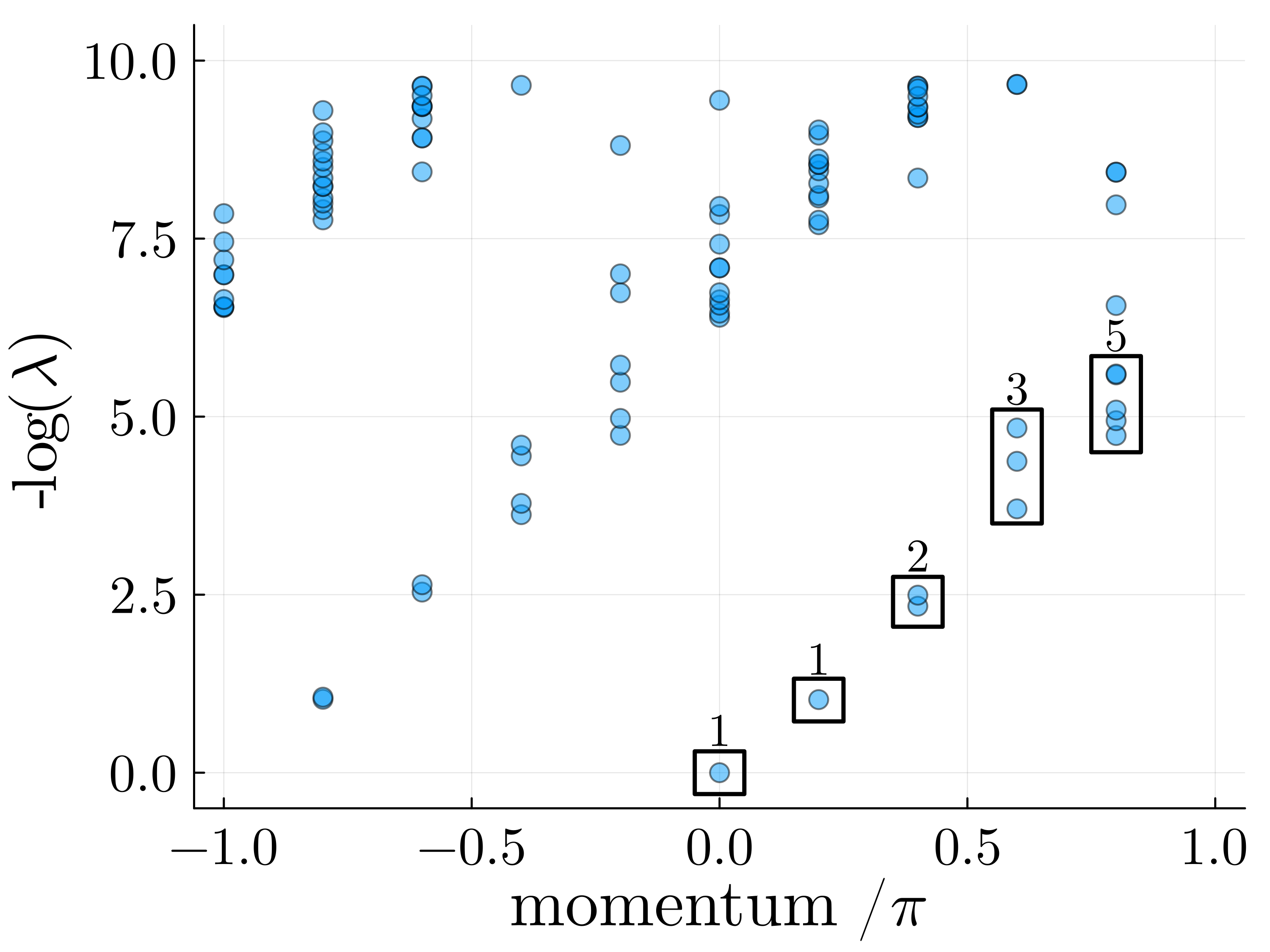}
    \caption{\raggedright \small Soft-core boson occupation between the Mott lobes for $\frac{U}{t} = 50$. We see again the plateau corresponding to the $\nu =  \frac{1}{2}$ Laughlin state and the corresponding hole analog but no additional plateaus at higher filling.}
    \label{fig:occ_softcore_pi_half}
\end{figure}

In addition to the calculation of the entanglement spectrum in the thermodynamic limit, we can gain information about the nature of the boundary modes by considering a system on a cylinder with finite circumference.
The discretization of the available momenta makes the counting of spectral degeneracies feasible: indeed, depending on what exactly is the nature of the boundary mode, we expect different counting patterns, e.g., (1, 1, 2, 3, 5, \ldots) for a chiral boson -- corresponding to the integer partitions of the added momentum in discrete units \cite{ManyBodyQFTBookXGW}.

Technically, this can be done by using Lanczos methods to look for the largest eigenvalues of the $\rho_l$ on a cylinder. Alternatively, one can also use again the calculation of the leading eigenvector in combination with the calculation of the excitations of periodic systems using open boundary MPS \cite{vandamme21ringsspectrum}. Once the leading eigenvalues and eigenvectors are determined with one of the methods above, one can use the periodic translation operator to identify the momentum associated with the different states. 
In the main text we showed the lowest branch of the entanglement spectrum. Here we show a larger window, where we can see an additional branch at negative momenta. This branch is actually comprised of two quasi-degenerate branches, for one additional or one missing particle in the system: Thus we observe a doubled counting, i.e., (2, 2, 4, 6, 10, ...).
Higher branches exhibit more complicated countings and are affected by numerical spurious effects.

Lastly, it should be mentioned the even for the calculation of the entanglement spectrum on a cylinder, we use the environment tensors from the CTMRG calculated on a plane, which is technically an additional approximation.

\section{Occupation plateaus for softcore-bosons at $\phi = \pi / 2$}
\begin{figure}[t]
    \centering
    \includegraphics[width=0.49\textwidth]{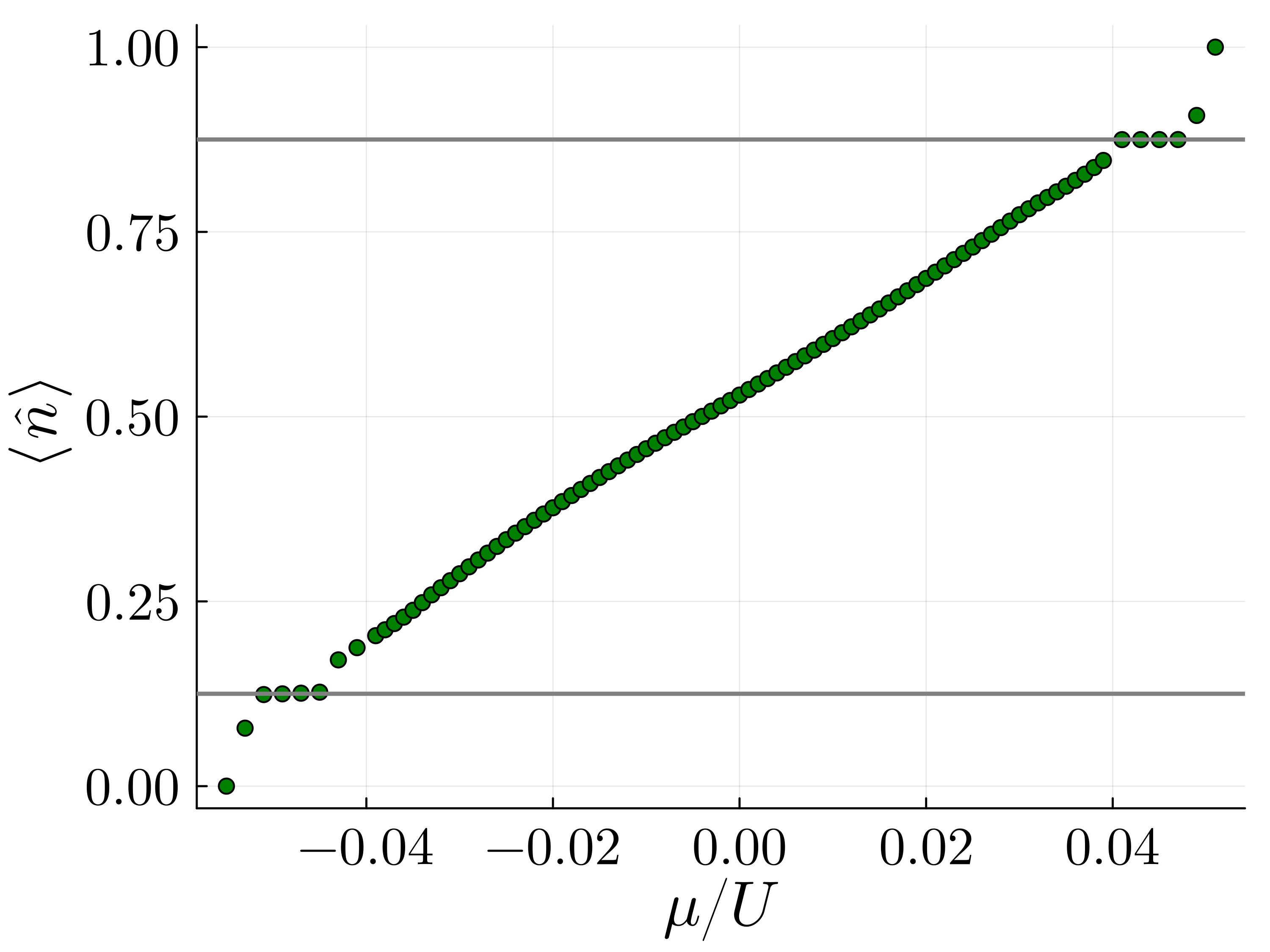}
    \caption{\raggedright \small Soft-core boson occupation between the Mott lobes for $\frac{U}{t} = 50$. We see again the plateau corresponding to the $\nu =  \frac{1}{2}$ Laughlin state and the corresponding hole analog but no additional plateaus at higher filling.}
    \label{fig:occ_softcore_pi_half}
\end{figure}

When investigating the bosonic Harper-Hofstadter model we investigated the case of soft-core bosons in the case of strong interactions $U/t = 50$, allowing for at most 2 bosons per site. The average occupation per site is shown in fig. \ref{fig:occ_softcore_pi_half}. We find a situation close to the one of hard-core bosons, which is not enirely unexpected at very strong local two-body interactions. Again, a incompressible plateau emerges at low and high filling corresponding to the Laughlin states of particles and holes respectively. We notice that, while the exact particle hole transformation presented in our discussion of the hard-core case does no longer hold here, it is reasonable to expect that the arguments extend qualitatively to the case of large finite interactions.

\section{Notes on the convergence with chiral gapped states}

In order to reach convergence during the optimization, one may encounter instabilities when the ground state is a gapped chiral topological state. 
In this scenario one can usually get back to a stable optimization by increasing the environment bond dimension before the instability occurs. 
It is also helpful to use the most accurate projectors possible in this case.
The numerical problems can be attributed to the artificial tail in the correlation functions that develops, when treating such gapped chiral topological states. 
In order to include these artificial tails -- which are crucial for representing such states with PEPS -- in the calculation of the energy, one needs a large environment bond dimensions. This also leads to a situation in which the CTMRG procedure need a large number of steps to converge. 
This is the main reason why the optimization of these states -- especially in the case of soft-core bosons -- can be challenging in practice as oftentimes a large number of optimization steps (on the order of hundreds to thousands) are necessary at quite large environment bond dimensions.

\end{document}